\documentclass[12pt,fleqn]{article}
\usepackage{graphicx}

\usepackage{latexsym}

\usepackage{amsmath}
\usepackage{amsthm}
\usepackage{amssymb}
\usepackage{amsfonts}

\begin{document}

\newcommand{\rf}[1]{(\ref{#1})}
\newcommand{\rff}[2]{(\ref{#1}\ref{#2})}

\newcommand{\ba}{\begin{array}}
\newcommand{\ea}{\end{array}}

\newcommand{\be}{\begin{equation}}
\newcommand{\ee}{\end{equation}}

\newcommand{\const}{{\rm const}}
\newcommand{\ep}{\varepsilon}
\newcommand{\Cl}{{\cal C}}
\newcommand{\rr}{\vec r}
\newcommand{\ph}{\varphi}
\newcommand{\R}{\mathbb R}
\newcommand{\N}{\mathbb N}

\newcommand{\e}{{\bf e}}

\newcommand{\m}{\left( \ba{r}}
\newcommand{\ema}{\ea \right)}
\newcommand{\mm}{\left( \ba{cc}}
\newcommand{\miv}{\left( \ba{cccc}}

\newcommand{\scal}[2]{\mbox{$\langle #1 \! \mid #2 \rangle $}}
\newcommand{\ods}{\par \vspace{0.5cm} \par}
\newcommand{\dis}{\displaystyle }
\newcommand{\mc}{\multicolumn}

\newtheorem{prop}{Proposition}
\newtheorem{Th}{Theorem}
\newtheorem{lem}{Lemma}
\newtheorem{rem}{Remark}
\newtheorem{cor}{Corollary}
\newtheorem{Def}{Definition}
\newtheorem{open}{Open problem}
\newtheorem{ex}{Example}
\newtheorem{exer}{Exercise}

\title{\bf Long-time behaviour of discretizations of the simple pendulum equation}

\author{
 {\bf Jan L.\ Cie\'sli\'nski}\thanks{\footnotesize
 e-mail: \tt janek\,@\,alpha.uwb.edu.pl}
\\ {\footnotesize Uniwersytet w Bia\l ymstoku,
Wydzia\l \ Fizyki, ul.\ Lipowa 41, 15-424
Bia\l ystok, Poland}
\\ {\bf Bogus\l aw Ratkiewicz}\thanks{\footnotesize
e-mail: \tt bograt\,@\,poczta.onet.pl}
\\ {\footnotesize I Liceum Og\'olnokszta\l c\c ace,
ul.\ \'Sr\'odmie\'scie 31, 16-300 August\'ow, Poland;}
\\ {\footnotesize Doctoral Studies, Wydzia\l \ Fizyki, Uniwersytet Adama Mickiewicza, Pozna\'n, Poland}
}

\date{}

\maketitle

\begin{abstract}
We compare the performance of several discretizations of simple pendulum
equation in a series of numerical experiments. The stress is put on the long-time
behaviour. We choose for the comparison numerical schemes which
preserve the qualitative features of solutions (like periodicity). All these schemes
are either symplectic maps or integrable (preserving the energy integral)  maps, or both.
We describe and explain systematic errors (produced by any method) in numerical computations 
of the period and the amplitude of oscillations.
We propose a new numerical scheme which is a modification of the discrete gradient method.
This discretization preserves (almost exactly) the period of small oscillations for any
time step. 
\end{abstract}

\ods


\section{Introduction}

New but more and more important direction in the numerical analysis is
geometric numerical integration \cite{HLW,Is,IZ,MQ}.
Numerical methods within this approach are
tailored for specific equations rather than for large general classes of equations.
The aim is to preserve qualitative features, invariants and geometric properties
of studied equations, e.g.,
integrals of motion, long-time behaviour and sometimes even trajectories
(but it is difficult, sometimes even impossible, to preserve all properties
by a single numerical scheme).
``Although the apparent desirability of this practice might be obvious at first glance,
it nonetheless calls for a justification'' \cite{Is-multi}.

In this paper we perform a series of numerical experiments
comparing the performance of several standard and geometric
methods on the example of the simple pendulum equation. The
equation itself is very well known but its discrete counterparts
show many interesting and unexpected features, for instance the
appearance of chaotic behaviour for large time steps \cite{FA,Yo}.
We focus our attention on the stability and time step dependence
of the period and the amplitude for several discretizations of the
simple pendulum (assuming that the time step is sufficiently small).  
We describe and explain small periodic oscillations of
the period and of the amplitude around their average values.

We confine our studies either to symplectic maps or to energy-preserving maps. It is well known that symplectic integrators
are very stable as far as the conservation of the energy is concerned. Since the beginning of 1990s they are
successfully used in the long time integration of the solar system \cite{SW,WH1,Yo}, see also \cite{BFRB,GBB}.
The reason is that using any symplectic scheme of $n$th order
the error of the Hamiltonian for an exponentially long time is of the order $O(\ep^n)$ where $\ep$ is the constant step
of the integration \cite{BG,HLW,MPQ}.
Therefore, in studies of the long-time behaviour, symplectic
algorithms have a great advantage at the very beginning. Fortunatelly, the class of symplectic integrators includes
such well known and relatively simple numerical schemes as the standard leap-frog method and the implicit midpoint rule.
In this paper we compare these classical methods with new geometric methods which preserve the energy integral.

We also propose a new discretization (a modification of the discrete gradient method)
which has some advantages: it is almost exact for small oscillations (even for large
time steps) and keeps some outstanding properties of the discrete gradient method 
(e.g., its precision in describing motions in the neighbourhood of
the separatrix).

\section{Symplectic discretizations of Newton equations}

We consider scalar autonomous Newton equations:
\be  \label{Newton}
\ddot \ph = f (\ph) \ ,
\ee
which can be written as the following first order system
\be   \label{simham}
\dot \ph = p \ ,  \quad
\dot p = f (\ph) \ .
 \ee
The equations are integrable for any function $f = f (\ph)$ (in this case by integrability we mean 
the existence of the integral of motion, compare \cite{Suris}). The energy conservation law reads
\be  \label{ener-int}
\frac{1}{2} {\dot \ph}^2 + V(\ph) = E \ , \qquad f (\ph ) = - \frac{d V (\ph)}{d \ph} \ , 
\ee
where $E = \const$. The Hamiltonian is given by
\be \label{H-V}
  H (p, q) = \frac{p^2}{2} + V (q) \ .
\ee
As an example to test quantitatively various numerical methods we will use
the simple pendulum equation
\be  \label{pendul}
  \ddot \ph = - k  \sin\ph \ .
\ee
In this case the energy conservation law has the form
\be  \label{pend1}
  \frac{1}{2} p^2 - k \cos\ph =  E \ .
\ee
The constant $k$ is not important. It can be eliminated by a change of the variable $t$.
In the sequel (in any numerical computations) we assume $k =1$.

By the discretization of \rf{Newton} we mean an $\ep$-family of difference equations (of the second order) 
which in the continuum limit $\ep \rightarrow 0$ yields \rf{Newton}. The initial conditions should be 
discretized as well, i.e., we have to map $\ph (0) \mapsto \ph_0$, $\dot \ph (0) \mapsto p_0$.

It is convenient to discretize \rf{simham} which automatically gives the discretization of $p$. 
Thus we have an $\ep$-dependent map $(\ph_n, p_n) \mapsto (\ph_{n+1}, p_{n+1})$. This map is called symplectic if  for any $n$ 
\be
d \ph_{n+1} \wedge d p_{n+1} = d \ph_n \wedge d p_n \ .
\ee
The following lemmas give a convenient characterization of symplectic maps and we will apply them in the next sections.

\begin{lem}  \label{PR}
The map \ $(\ph_n, p_n) \mapsto (\ph_{n+1}, p_{n+1})$,  
implicitly defined by
\be  \label{mFG}
  \ph_{n+1} - \ph_n = P (p_n, p_{n+1}, \ep) \ , \quad p_{n+1} - p_n = R (\ph_n, \ph_{n+1}, \ep) \ ,
\ee
where $P$ and $R$ are differentiable functions, is symplectic if and only if  
\be
\frac{\partial P}{\partial p_n} \frac{\partial R}{\partial \ph_n} = 
\frac{\partial P}{\partial p_{n+1}} \frac{\partial R}{\partial \ph_{n+1}} \neq 1 \ .
\ee  
\end{lem}

\noindent The proof is straightforward. Differentiating \rf{mFG} we get
\[  \ba{l}
d \ph_{n+1} - d \ph_n = P,_1 d p_n + P,_2 d p_{n+1} \ , \\
d p_{n+1} - d p_n = R,_1 d \ph_n + R,_2 d \ph_{n+1} \ ,
\ea \]
(where the comma denotes partial differentiation). Then 
\[  \ba{l} \dis
d \ph_{n+1} = \frac{1 + P,_2 R,_1}{1 - P,_2 R,_2} \ d \ph_n + \frac{P,_1 + P,_2}{1 - P,_2 R,_2} \ d p_n \ , \\[2ex] 
\dis 
d p_{n+1} = \frac{R,_1 + R,_2}{1 - P,_2 R,_2} \ d \ph_n + \frac{1 + P,_1 R,_2}{1 - P,_2 R,_2} \ d p_n \ , 
\ea \]
provided that $P,_2 R,_2 \neq 1$ (this condition means that the map defined by $P, R$ 
is non-degenerate). Therefore  
\[  \dis
d \ph_{n+1} \wedge d p_{n+1} = \frac{1 - P,_1 R,_1}{1 - P,_2 R,_2} \ d \ph_n \wedge d p_n \ .
\]
Hence the map is symplectic if $P,_1 R,_1 = P,_2 R,_2 \neq 1$ which ends the proof.  

\begin{lem}  \label{AB}
The map \ $(\ph_n, p_n) \mapsto (\ph_{n+1}, p_{n+1})$,  
defined by
\be  \label{mAB}
  \ph_{n+1} - A (\ph_n, \ep) + \ph_{n-1} = 0 \ , \quad p_n = \mu_0 (\ep) \ \ph_{n+1} + B (\ph_n, \ep) \ ,
\ee
is symplectic for any differentiable functions $A, B$.
\end{lem}

\noindent  In order to prove Lemma~\ref{AB} we compute
\[
d p_{n+1} = \mu_0 \ d \ph_{n+2} + TB' d \ph_{n+1} = \mu_0 \ TA' \ d \ph_{n+1} - \mu_0 \ d \ph_n + TB' \ d \ph_{n+1} \ , 
\]
where the prime denotes the differentiation and $T$ denotes the shift. Therefore
\[
d \ph_{n+1} \wedge d p_{n+1} = - \mu_0 \ d \ph_{n+1} \wedge d \ph_n \ .
\]
On the other hand \ $d \ph_n \wedge d p_n = \mu_0 \ d \ph_n \wedge d \ph_{n+1}$, which ends the proof.

\section{Nonintegrable symplectic discretizations}
\label{sec-stand}

In this section we present some well known discretizations which preserve the symplectic structure of the
Newton equations (compare \cite{HLW}, p. 189-190) but have no integrals of motion.

\subsection{Standard discretization}

The standard discretization of the simple pendulum equation
\be  \label{standard}
\frac{\ph_{n+1} - 2 \ph_{n} + \ph_{n-1}}{\ep^2} = - k   \sin \ph_n
\ee
is non-integrable \cite{Suris}. This discretization can be obtained by the application of
either leap-frog (St\"ormer-Verlet) scheme or one of the symplectic splitting methods.
It is interesting that we get the same discrete equation \rf{standard} but a different dependence
of $p_n$ on $\ph_n, \ph_{n+1}$ (compare \rf{PSV}, \rf{PSS}):
\be \label{pedy3}
p_n = \frac{\ph_{n+1}- \ph_n}{\ep} + c k \ep \sin \ph_n \ ,
\ee
where $c = 0, \frac{1}{2}, 1$. By virue of Lemma~\ref{AB} standard discretizations are symplectic 
(for any $c$).

\subsection{St\"ormer-Verlet (leap-frog) scheme}

The numerical integration scheme
\be \left\{ \ba{l}  \label{SVzaba}
p_{n+\frac{1}{2} } = p_n + \frac{1}{2} \ep f (\ph_n) \ , \\[1ex]
\ph_{n+1} = \ph_n + \ep p_{n+\frac{1}{2} } \ , \\[1ex]
p_{n+1} = p_{n+\frac{1}{2} }+ \frac{1}{2} \ep f (\ph_{n+1}) \ ,
 \ea \right.
\ee
is known as the St\"ormer-Verlet (or leap-frog) method (compare, e.g., \cite{HLW}).
Eliminating $p_{n + \frac{1}{2}}$, we can easily formulate the St\"ormer-Verlet as a one-step method:
\be \ba{l}   \label{SVonestep}
\ph_{n+1} = \ph_n + \ep p_n + \frac{1}{2} \ep^2 f (\ph_n) \ , \\[3ex]
p_{n+1} = p_n + \frac{1}{2} \ep \left( f (\ph_n) + f \left( \ph_n + \ep p_n + \frac{1}{2} \ep^2 f (\ph_n)  \right) \right) \ .
\ea \ee
We can also formulate this method as
\be  \label{stanf}
\frac{\ph_{n+1} - 2 \ph_n + \ph_{n-1}}{\ep^2} = f (\ph_n) \ ,
\ee
\be \label{PSV}
p_n = \frac{\ph_{n+1}- \ph_n}{\ep} - \frac{\ep}{2} f (\ph_n) \ .
\ee
In the simple pendulum case ($f (\ph) = - k \sin\ph$) we recognize in
the equations \rf{stanf}, \rf{PSV}
the standard discretization \rf{standard}, \rf{pedy3} with $c = 1/2$.

\subsection{Symplectic splitting methods}

The system \rf{simham} belongs to the class of
``partitioned systems'' which have the form
\be
\dot \ph = g (\ph, p) \ , \quad \dot p = h (\ph, p) \ ,
\ee
where $g, h$ are given functions of two variables. We can discretize such systems
in one of the following two ways:
\be
\ph_{n+1} = \ph_n + \ep g (\ph_n, p_{n+1} ) \ , \quad p_{n+1} = p_n + \ep h (\ph_n, p_{n+1} ) \ ,
\ee
\be
\ph_{n+1} = \ph_n + \ep g (\ph_{n+1}, p_n ) \ , \quad p_{n+1} = p_n + \ep h (\ph_{n+1}, p_n ) \ .
\ee
Both these discretizations are called either symplectic Euler methods \cite{HLW} or symplectic splitting methods \cite{MQR2}.
In our case (see \rf{simham}) we have, respectively,
\be   \label{ss1}
\ph_{n+1} = \ph_n + \ep  p_{n+1} \ , \quad
p_{n+1} = p_n + \ep f (\ph_n ) \ ,
\ee
\be \label{ss0}
\ph_{n+1} = \ph_n + \ep  p_n \ , \quad
p_{n+1} = p_n + \ep f ( \ph_{n+1} ) \ .
\ee
Finally, both \rf{ss1} and \rf{ss0} yield \rf{stanf}, but
instead of \rf{PSV} we have
\be \label{PSS}
p_n = \frac{\ph_{n+1}- \ph_n}{\ep} - \ep f (\ph_n) \quad {\rm or} \quad p_n = \frac{\ph_{n+1}- \ph_n}{\ep} \ ,
\ee
i.e., in the simple pendulum case we get \rf{pedy3} with $c=1$ and $c=0$, respectively.

\subsection{Implicit midpoint rule}

Any first order equation $\dot x = F (x)$  can be discretized using implicit midpoint rule
(which coincides with the implicit 1-stage Gauss-Legendre-Runge-Kutta method, compare \cite{HLW}).
The first derivative is replaced by the difference quotient and the right hand side is evaluated
at midpoint $\frac{1}{2} (x_n + x_{n+1})$. In the case of the simplest Hamiltonian systems, given by \rf{simham}, 
we have:
\be \ba{l}  \label{mid1}
\ph_{k+1} = \ph_k + \frac{1}{2} \ep ( p_{k} + p_{k+1} ) \ , \\[2ex]
p_{k+1} = p_k + \ep f (\frac{\ph_k + \ph_{k+1}}{2} ) \ .
\ea \ee
In the special case of the simple pendulum we get
\be \ba{l}
\frac{\ph_{k+1} - 2 \ph_k + \ph_{k-1}}{\ep^2} = - \frac{1}{2} k \left( \sin (\frac{\ph_{k+1}+ \ph_k}{2} )
 + \sin \left( \frac{\ph_k + \ph_{k-1} }{2} \right) \right)  \ , \\[2ex]
p_k = \frac{\ph_{k+1} - \ph_k}{\ep} + \frac{1}{2} \ep k \sin \frac{\ph_{k+1} + \ph_k}{2}  \ .
\ea \ee
The implicit midpoint rule has quite good properties: this is a symplectic, time-reversible method of order 2. 
The symplecticity follows directly from Lemma~\ref{PR}. Indeed, \rf{mid1} implies $P,_1 = P,_2$ and $R,_1 = R,_2$.

\section{Projection methods}

Non-integrable discretizations can be modified so as to preserve the energy integral "by force", i.e., projecting
the result of every step on the constant energy manifold. In principle, any one-step method can be converted into
the corresponding projection method. In this paper we apply these procedures 
to the St\"ormer-Verlet (leap-frog) method. 
 Therefore referring to the "standard projection" 
and "symmetric projection" we always mean standard (or symmetric) projection applied to the leap-frog scheme. 

\subsection{Standard projection method}

There are given a first order equation $\dot x = F (x)$, $x \in {\mathbb R}^2$, any one-step
numerical method $x_{n+1} = \Phi_\ep (x_n)$
(a discretization of the ODE), and a constraint $ g (x) = 0$ \  we would like to preserve.
The standard projection consists in computing $\tilde x_{n+1} := \Phi_\ep (x_n)$, and then
orthogonally projecting $\tilde x_{n+1}$ on the manifold $g (x) = 0$, see \cite{HLW}.
This projection, denoted by $x_{n+1}$, yields the next step: $x_n \rightarrow x_{n+1}$. In other words, we define
\be  \label{iter1}
x_{n+1} = \tilde x_{n+1}  + \lambda   \nabla g (\tilde x_{n+1} )
\ee
where $\lambda$ is such that $g (x_{n+1}) = 0$.

Applying this approach to the simple pendulum \rf{pendul} it is convenient to define $x$ as
\be
    x = \left( \ph, \ \frac{\dot \ph}{\omega} \right) \equiv (\ph,\ p) \ ,
\ee
where $\omega = \sqrt{k}$. The above definition of $p$ yields dimensionless components of $x$.
If $k=1$ (which is assumed throughout this paper), then this definition of $p$ coincides
with the previous one, see \rf{simham}. The constraint $g(x)=0$ is given by \rf{pend1}, i.e.,
\be
   g (x) =  \frac{1}{2}  p^2 - \cos\ph   - h \ .
\ee
where $h = E/\omega^2 $.
The equation \rf{iter1} becomes
\be
 \ph_{n+1} = \tilde\ph_{n+1} + \lambda \sin\tilde\ph_{n+1} \ , \quad
p_{n+1} = (1 + \lambda) \tilde p_{n+1}
\ee
and $\lambda$ is computed from
\be  \label{constr2}
\frac{1}{2} (1 + \lambda)^2 \tilde p_{n+1}^2 - \cos (\tilde\ph_n + \lambda \sin\tilde\ph_{n+1} ) = h \ .
\ee 
In order to solve \rf{constr2} 
we use Newton's iteration $\lambda_{j+1} = \lambda_j - \Delta \lambda_j $, where
\be
\Delta \lambda_j = - \frac{\frac{1}{2} (1 + \lambda_j)^2 \tilde p_{n+1}^2 - \cos (\tilde\ph_n +
\lambda_j \sin\tilde\ph_{n+1} ) - h }{{{\tilde p}_{n+1}^2} + \sin^2\tilde\ph_{n+1}} \ ,
\ee 
and it is sufficient and convenient to choose $\lambda_0 = 0$. 
The approximated solution to \rf{constr2} is given by  $\lambda = \lim_{j\rightarrow \infty} \lambda_j$.

\subsection{Symmetric projection method}

A one-step algorithm $x_{n+1} = \Phi_\ep (x_n)$ is called symmetric (or time-reversible) if $\Phi_{-\ep} = \Phi_\ep^{-1}$.
Equations of the classical mechanics are time-reversible, therefore the preservation of this property is convenient
and is expected to improve numerical results. The symplectic splitting methods
are not time-reversible while the St\"ormer-Verlet method and implicit midpoint rule are symmetric.
The symmetry can be easily noticed in the form \rf{SVzaba} of the leap-frog method.

The symmetric projection method preserves the time-reversibility. The method is applied under similar assumptions as
standard projection (additionally we demand the time-reversibility of $\Phi_\ep$) and consists of the following steps
\cite{AR,H-sym}:
\be  \ba{l}
 \hat x_n = x_n + \lambda \nabla g (x_n)  \ , \\[2ex]
\tilde x_{n+1} = \Phi_\ep (\hat x_n) \ , \\[2ex]
x_{n+1} = \tilde x_{n+1} + \lambda \nabla g (x_{n+1}) \ ,
\ea \ee
where we assume $g (x_n) = 0$ and compute
the parameter $\lambda$ from the condition $g (x_{n+1} ) = 0$.

\section{Integrable discretizations}

Throughout this paper by integrability we mean the existence of an integral of motion.
The Newton equation \rf{Newton} has the energy integral \rf{ener-int}. Its discretization
is called integrable when it has an integral of motion as well. In the continuum limit this integral
becomes the energy integral, so it may be treated as a discrete analogue of the energy.

\subsection{Standard-like discretizations}

Standard-like discretizations are defined by \cite{Suris}
\be  \ba{l} \label{stlike}
\ph_{n+1} = \ph_n + \ep p_{n+1} \ , \\[2ex]
p_{n+1} = p_n + \ep F (\ph_n,\ep)
\ea \ee
where $F$ has to satisfy $F (\ph_n, 0) = f (\ph_n)$.
 For a given $f$ there exist inifinitely many functions $F$ satisfying this conditions. 
All of them are symplectic, which can be easily seen applying Lemma~\ref{PR} with $P,_1 = R,_2 = 0$. 
Similarly as in Section~\ref{sec-stand} we obtain from \rf{stlike}:
\be  \ba{l} \label{Fqn}
\ph_{n+1} -  2 \ph_n + \ph_{n-1} = \ep^2 F (\ph_n, \ep) \ ,
\\[3ex] \displaystyle
p_n = \frac{\ph_{n+1} - \ph_n}{\ep} - \ep F (\ph_n, \ep) = \frac{\ph_n - \ph_{n-1}}{\ep} \ .
\ea \ee

We are interested in integrable cases, i.e., in discretizations preserving the energy integral.
Suris found that two standard-like discretization of the simple pendulum are integrable \cite{Sur89,Suris}:
\be  \label{Sur1}
\ph_{n+1} - 2 \ph_n + \ph_{n-1} = - 2 \arctan \left( \frac{ k \ep^2 \sin\ph_n}{2 + k \ep^2 \cos\ph_n} \right) \ ,
\ee
\be   \label{Sur2}
\ph_{n+1} - 2 \ph_n + \ph_{n-1} = - 4 \arctan \left( \frac{ k \ep^2 \sin\ph_n}{4 + k \ep^2 \cos\ph_n} \right) \ .
\ee
The equation \rf{Sur1}, referred to as the Suris1 scheme, has the integral of motion given by
\be  \label{E1-Sur1}
E_1 =  \frac{1}{2} \left( \frac{2 \sin\frac{\ph_{n+1}-\ph_n}{2}}{ \ep } \right)^2 - \frac{1}{2}  k \left( \cos\ph_n + \cos\ph_{n+1} \right)  \ ,
\ee
or, in terms of $\ph_n$ and $p_n$,
\be  \label{E-Sur1}
E_1 = \frac{1 - \cos\ep p_n }{\ep^2} - \frac{1}{2} k \left( \cos\ph_n + \cos (\ph_n - \ep p_n) \right) \ .
\ee
The equation \rf{Sur2}, referred to as the Suris2 scheme, has the following integral of motion
\be    \label{E2-Sur2}
   E_2 =  \frac{1}{2} \left(  \frac{4 \sin\frac{\ph_{n+1}-\ph_n}{4}}{ \ep } \right)^2 - k  \cos { \frac{ \ph_n + \ph_{n+1} }{2} }  \ ,
\ee
which can be expressed in terms of $\ph_n$ and $p_n$ as follows
\be \label{E-Sur2}
 E_2 = \frac{4}{\ep^2} \left( 1  - \cos \frac{\ep p_n}{2} \right) - k \cos ( \ph_n - \frac{\ep p_n}{2}  ) \ .
\ee
One can verify the preservation of these integrals by direct computation.

\subsection{Discrete gradient method}

The discrete gradient method \cite{MQR2,QC,QT} is a general and very powerful method to
generate numerical schemes preserving any number of integrals of motion and some other properties
\cite{MQR1}. However, this method in general is not symplectic. 
In this paper we need to preserve one integral (the energy) and the system is hamiltonian, compare \rf{H-V},  
\be   \label{hameq}
  \dot \ph = \frac{\partial H}{\partial p} \ , \quad
\dot p = - \frac{\partial H}{\partial \ph} \ .
\ee
In such case  the discrete gradient method reduces to the following simple scheme.
Left hand sides of the formulas \rf{hameq} are discretized in the simplest way (difference quotients) while the right hand sides
are replaced by the so called discrete (or average) gradients:
\be
\frac{\ph_{n+1} - \ph_n}{\ep} = \frac{\Delta H}{\Delta \ph}\ , \quad
\frac{p_{n+1} - p_n }{\ep} = - \frac{\Delta H}{\Delta p} \ .
\ee
The discrete gradient $\bar \nabla H \equiv \left( \frac{\Delta H}{\Delta \ph} , \ \frac{\Delta H}{\Delta p} \right)$
of a differentiable function $H ( \ph, p)$ by definition (see \cite{MQR2}) satisfies the condition
\be
  H (\ph_{n+1}, p_{n+1}) - H (\ph_n, p_n) = \frac{\Delta H}{\Delta \ph} ( \ph_{n+1} - \ph_n ) +
\frac{\Delta H}{\Delta p} ( p_{n+1} - p_n ) \ .
\ee
The explicit form of $\bar \nabla H$  is, in general, not unique. One of the possibilities is the coordinate increment
discrete gradient \cite{IA}
\be
  \frac{\Delta H}{\Delta \ph}  = \frac{ H ( \ph_{n+1}, p_n ) - H (\ph_n, p_n) }{\ph_{n+1} - \ph_n } \ ,
\quad \frac{\Delta H}{\Delta p}  = \frac{ H (\ph_{n+1}, p_{n+1} ) - H ( \ph_{n+1}, p_n) }{ p_{n+1} - p_n } \ .
\ee
Other possibilities are, for instance, mean value discrete gradient \cite{MQR2} and midpoint discrete gradient \cite{Gon}.
All these definitions coincide in the case $H (\ph, p) = T (p) + V (\ph)$. In such case $\bar \nabla H = \bar \nabla T + \bar \nabla V$, where
\be
\bar \nabla T = \frac{ T (p_{n+1}) - T (p_n )}{p_{n+1} - p_n} \ , \quad
\bar \nabla V = \frac{ V (\ph_{n+1}) - V (\ph_n )}{\ph_{n+1} - \ph_n} \ .
\ee
Thus we have got the discrete gradient scheme:
\be \left\{ \ba{l} \displaystyle   \label{disgrad}
 \frac{p_{n+1} + p_n }{2} = \frac{ \ph_{n+1} - \ph_n}{\ep}   \ , \\[2ex]
\displaystyle
\frac{ p_{n+1} - p_n }{\ep} = - \frac{ V (\ph_{n+1}) - V(\ph_n) }{\ph_{n+1} - \ph_n} \ . \\[1ex]
 \ea \right.
\ee
This numerical scheme can be also obtained as a special case of the modified midpoint rule 
\cite{LG}. 
The system \rf{disgrad} can be rewritten
as the following second order equation for $\ph_n$ plus the defining equation for $p_n$:
\be \ba{l}  \label{pQ} \dis
\frac{\ph_{n+1} - 2 \ph_n + \ph_{n-1} }{\ep^2} =  - \frac{1}{2} \left( \frac{ V (\ph_{n+1}) - V (\ph_n) }{  \ph_{n+1} - \ph_n  }
+ \frac{ V (\ph_n)  - V (\ph_{n-1}) }{  \ph_n - \ph_{n-1}  }  \right)    \\[4ex] \dis
p_n = \frac{ \ph_{n+1} - \ph_n}{\ep} + \frac{1}{2} \ep  \left(  \frac{ V (\ph_{n+1}) - V (\ph_n )}{ \ph_{n+1} - \ph_n} \right)  \ .
\ea \ee
Substituting  $ V(\ph) = - k \cos\ph$ we get the simple pendulum case. Multiplying both equations 
\rf{disgrad} side by side, we easily prove that the system \rf{pQ} has the first integral
\be
    E = \frac{1}{2} p_n^2 + V (\ph_n )
\ee
which exactly coincides with the hamiltonian \rf{H-V} evaluated at $\ph_n$, $p_n$. Note that 
the integrals of motion
\rf{E-Sur1}, \rf{E-Sur2} coincide with \rf{H-V} (where $V (\ph) = - k \cos\ph$) only approximately,
in the limit $\ep \rightarrow 0$.

\section{A correction which preserves the period of small oscillations}

The classical harmonic oscillator equation $\ddot \ph + \omega^2 \ph = 0$ admits the exact discretization
(\cite{CR}, compare also \cite{Ag,Reid}), i.e., a discretization such that the solution $\ph(t)$ evaluated at
$n \ep$ equals $\ph_n$ (for any $\ep$, and any $n$):
\be \ba{l}  \label{exact-osc}
 \ph_{n+1} - 2 \ph_n \cos\ep\omega + \ph_{n-1} = 0  , \\[2ex]
 \dis p_n = \frac{\omega}{\sin\omega\ep} \left( \ph_{n+1} - \ph_n \cos\omega\ep  \right) \ .
\ea \ee
The energy is also exactly preserved, i.e.,
\be
 E = \frac{1}{2} p_n^2 + \frac{1}{2} \omega^2 \ph_n^2
\ee
does not depend  on $n$ (which can be easily checked by direct calculation). The existence of the exact discretization
of the harmonic oscillator equation has been recently used to discretize the Kepler problem (preserving all integrals
of motion and trajectories) \cite{Ci-Kep}.

We consider the class of Newton equations \rf{Newton}. Let us confine ourselves to equations
which have a stable equilibrium  at $\ph = 0$, i.e., $f'(0) < 0$. Then
$V = V(\ph)$ has a local minimum at $\ph = 0$, i.e., $V'(0) = f (0) = 0$. We denote
\be \label{omega0}
  \omega_0 = \sqrt{V''(0)} \ .
\ee
Thus
\be \label{Tay}
  V (\ph) = V_0 + \frac{1}{2} \omega_0^2 \ph^2 + \ldots \ ,
\ee
and small oscillations around the equilibrium  can be approximated by the classical harmonic
oscillator equation with $\omega = \omega_0$.

Do exist  discretizations which in the limit $\ph_n \approx 0$ ($\ep$ is fixed) become exact? Known discretizations,
including those presented in this paper, do not have this property. Fortunatelly, we found such discretization
by modifying the discrete gradient method. It is sufficient to replace
$\ep$ by some function $\delta = \delta (\ep)$ in the formulae \rf{disgrad}. The form of this function
will be obtained by the comparison with the harmonic oscillator equation (in the limit $\ph \approx 0$).

We linearize the equations \rf{pQ} (with $\ep$ replaced by $\delta$) around $\ph_n = 0$
(i.e., we take into account \rf{Tay}). Thus we get
\be  \ba{l}
\dis \frac{\ph_{n+1} - 2 \ph_n + \ph_{n-1}}{\delta^2}  = - \frac{\omega_0^2}{4} \left( \ph_{n+1} + 2 \ph_n + \ph_{n-1} \right) \ ,  \\[3ex]
\dis p_n = \frac{ \ph_{n+1} - \ph_n}{\delta} + \frac{1}{4} \omega_0^2 \delta  \left(   \ph_{n+1} + \ph_n  \right)  \ ,
\ea \ee
which is equivalent to
\be \ba{l}  \label{exact-?}
\dis \ph_{n+1} - 2 \left( \frac{4 - \omega_0^2 \delta^2}{4 + \omega_0^2 \delta^2} \right) \ph_n + \ph_{n-1} = 0 \ , \\[3ex]
\dis p_n = \frac{4 + \omega_0^2 \delta^2}{4 \delta} \left( \ph_{n+1} - \left( \frac{4 - \omega_0^2 \delta^2}{4 + \omega_0^2 \delta^2} \right) \ph_n \right) \ .
\ea \ee
We compare \rf{exact-osc} with \rf{exact-?}. Both systems coincide if and only if
\be  \label{komega}
\frac{4 - \omega_0^2 \delta^2}{4 + \omega_0^2 \delta^2} = \cos\ep\omega \ , \quad \frac{4 + \omega_0^2 \delta^2}{4 \delta} = \frac{\omega}{\sin\ep\omega} \ .
\ee
Solving the system \rf{komega} we get
\be \label{delta}
\omega =\omega_0 \ , \quad  \delta =  \frac{2}{\omega_0} \tan \left( \frac{\ep \omega_0}{2} \right)  \ .
\ee
Therefore, we propose the following new discretization of the Newton equation \rf{Newton}, \rf{ener-int} 
(modified discrete gradient scheme):
\be  \label{dis-delta}
\ba{l} \dis
\frac{\ph_{n+1} - 2 \ph_n + \ph_{n-1} }{\delta^2} = - \frac{1}{2} \left( \frac{ V (\ph_{n+1})
- V(\ph_n) }{ \ph_{n+1} - \ph_n} + \frac{ V (\ph_n) - V (\ph_{n-1})}{\ph_n - \ph_{n-1}} \right)  \\[4ex] \dis
p_n = \frac{ \ph_{n+1} - \ph_n}{\delta} + \frac{1}{2} \delta \left( \frac{ V (\ph_{n+1}) - V (\ph_n) }{ \ph_{n+1} - \ph_n } \right)
\ea \ee
where $\delta$ is defined by \rf{delta} (and $\omega_0$ is given by \rf{omega0}). This discretization becomes exact for small
oscillations for any fixed $\ep$. It means that for $\ph_n \approx 0$ the period and the amplitude of the approximated solution should be very close to
the exact values (even for large $\ep$!). In the next sections we will verify  this point 
experimentally.

\section{Numerical experiments}

We performed a number of numerical experiments applying the numerical  
schemes presented above. 
The initial data were parameterized by the velocity
$p_0$ while the initial position was always the same: $\ph_0 = 0$. 
In the continuous case \rf{pendul} we have 3 possibilities: oscillating motion ($|p_0| < 2$), rotating 
motion ($|p_0| > 2$) and the motion along the separatrix ($p_0 = \pm 2$), from 
$\ph = 0$ to (asymptotically) $\ph = \pm \pi$.
The (theoretical) amplitude $A_{th}$ for the oscillating motions can be easily computed 
from the energy conservation law \rf{pend1} (where $k=1$, i.e., $\frac{1}{2} p_0^2 - 1 = -\cos A_{th}$):
\be
      2 \sin \frac{A_{th}}{2} = p_0 \ .
\ee
In particular, we performed many numerical computations for the following initial data: 
\begin{itemize}
\item  $p_0 = 0.1$, then $A_{th} \approx 0.0318443 \pi \approx 0.1000417$ (small amplitude)
\item  $p_0 = 1.8$, then $A_{th} \approx 0.712867 \pi \approx 2.239539$ (very large amplitude). 
\end{itemize}
To estimate the actual amplitude of a given discrete simulation  we
apply the following procedure: if $\ph_m$ is a local maximum of the discrete trajectory
 (i.e., $\ph_m > \ph_{m-1}$ and
$\ph_m > \ph_{m+1}$),
then we estimate the maximum of the approximated function by the maximum of the parabola best fitted to 
the following five points: $\ph_{m-2}, \ph_{m-1}, \ph_m, \ph_{m+1}, \ph_{m+2}$. 
The analogical procedure is done also at local minima (we take the absolute value of the obtained minimum). 
Thus we obtain a sequence of the amplitudes, $A_N$. The index
$N$ is common for all extrema (maxima and minima), and on some figures we denote 
it by $N_{1/2}$ (the number of half periods) to discern it from $N$ (the number of periods). 

Every numerical scheme used in the present paper yields a discrete trajectory with rather stable amplitude. 
It is not constant but oscillates in a regular way around an average value:  
\be
 A_N = A ( 1  + \alpha_N ) \ ,
\ee
where both the average amplitude $A$ and relative (dimensionless)
oscillations $\alpha_N$ can depend on the time step $\ep$ and on the initial
velocity $p_0$,
i.e., $A = A (p_0, \ep)$ and
$\alpha_N = \alpha_N (p_0, \ep) $. Of course, both $A$ and $\alpha_N$
differ for different numerical schemes.

In a similar way we estimated the period of discrete motions. 
The exact periodicity ($\ph_{k+n} = \ph_k$ for some $k, n$) is a rare phenomenon and, of course,
 we did not observe it.
To define the approximate period we fit a continuous curve to the discrete graph,
estimate zeros of this function, and compute the distance between the neighbouring zeros.

Suppose that $\ph_{m} \ph_{m+1} < 0$ \ for some $m$. It means that one of the zeros, say $z_N$, lays between
$\ph_m$ and $\ph_{m+1}$. We estimate it by
 zero of the interpolating cubic polynomial based on the points $\ph_{m-1}$, $\ph_{m}$,
$\ph_{m+1}$, $\ph_{m+2}$ (another natural, but less accurate, possibility could be a line joining 
$\ph_m$ and $\ph_{m+1}$).
Then, denoting subsequent estimated zeros by $z_N$ ($N=1,2,3,\ldots$) and
$z_0 = \ph_0 = 0$,
we define
\be
T_N = z_{2N} - z_{2N-2}  \ ,
\ee
which we take as an estimate of the period.

Our numerical experiments have shown that $T_N$ is not exactly constant but
oscillates with a relatively small amplitude. The average value of $T_N$
is constant with high accuracy (see the next section). 
Therefore we have
\be
 T_N = T  ( 1  + \tau_N ) \ ,
\ee
where both the average period $T$ and relative (dimensionless)
oscillations $\tau_N$ can depend on the time step $\ep$ and on the initial
velocity $p_0$,
i.e., $T = T (p_0, \ep)$ and
$\tau_N = \tau_N (p_0, \ep) $. Moreover, $T$ and $\tau_N$
essentially depend on the discretization (numerical scheme).

The amplitude of small oscillations is defined in a natural way
\be
\tau (\ep, p_0) := \max_N |\tau_N (\ep, p_0) | \ , \qquad \alpha (\ep, p_0) := \max_N |\alpha_N (\ep, p_0) \ .
\ee
Fortunatelly, $|\tau_N|$ and $|\alpha_N|$ oscillate (as functions of $N$), with small amplitudes, 
in a very regular way. Thus 
we can estimate $\tau (\ep, p_0)$ and $\alpha (\ep, p_0)$ considering a series of, say, 40 local extrema of 
$\tau_N$ and $\alpha_N$, and taking an average value.

\section{Periodicity and stability}

Discrete trajectories generated by symplectic or integrable schemes considered in our paper 
are stable for $\ep$ which are not too large (for very large $\ep$ one can observe 
chaotic behaviour, \cite{FA,Yo}). We confine ourselves to sufficiently small $\ep$, i.e. 
$\ep \leqslant 0.5$,  
but sometimes (for $p_0 < 1.5$) we can take even $\ep \approx 1$. In this region the motion is very stable 
and both the average period $T$ and the average amplitude $A$ are well defined. 
The average amplitude is computed simply as 
\be
 A_{avg}(N,M) = \frac{1}{M} \sum_{j=0}^{M-1}  |A_{N+j}| \ , 
\ee
where we usually assume $M = 50$. The definition of the average period is similar.  
In many cases we use the formula
\be  \label{TNM}
T_{avg} (N, M)  = \frac{1}{M} \left( z_{N+ 2 M} - z_N \right) \ ,  
\ee
where the dependence (very essential!) on $\ep$ and $p_0$ is omitted for the sake of brevity. 
Note that $T_N \equiv T_{avg} (2N-2, 1)$.  
Computing $T_{avg}$ it is necessary to choose $M$ arbitrarily, we usually take $M=20$. 
Sometimes we denote $N \equiv N_0$ to point out that the average is taken over 
indices greater than $N_0$. 

Considering very long discrete evolutions (many thousands of periods) we use  
another definition of the average period. Namely, 
we average $T_{avg} (N, M)$ over some range of the parameter $M$ ($K < M \leqslant L$):
\be
  {\bar T}_{avg} (N,K,L) = \frac{1}{L-K} \sum_{M = K+1}^{L}  T_{avg} (N,M) \ .
\ee
Usually we assume $K = 100$, $L = 200$. 

All discretizations considered in the present paper are characterized by very high stability 
of the period and the amplitude. 
One can hardly notice any dependence of $T_{avg}$ and $A_{avg}$ on $N$, even when 
testing very large $N$ 
(like $10^3$, $10^5$ or $10^6$), and ${\bar T}_{avg}$ is even more stable. 

\begin{figure}
\caption{\small $T_{avg} (N_0, 20)$ for the Suris1 scheme ($N_0 < 3100$), $\ep = 0.2$, $p_0 = 1.95$,  $T_{th} = 11,65758528$, 
$T = 11,88884005$. } \label{Suris1-195-02} \par
\includegraphics[height=0.35\textheight]{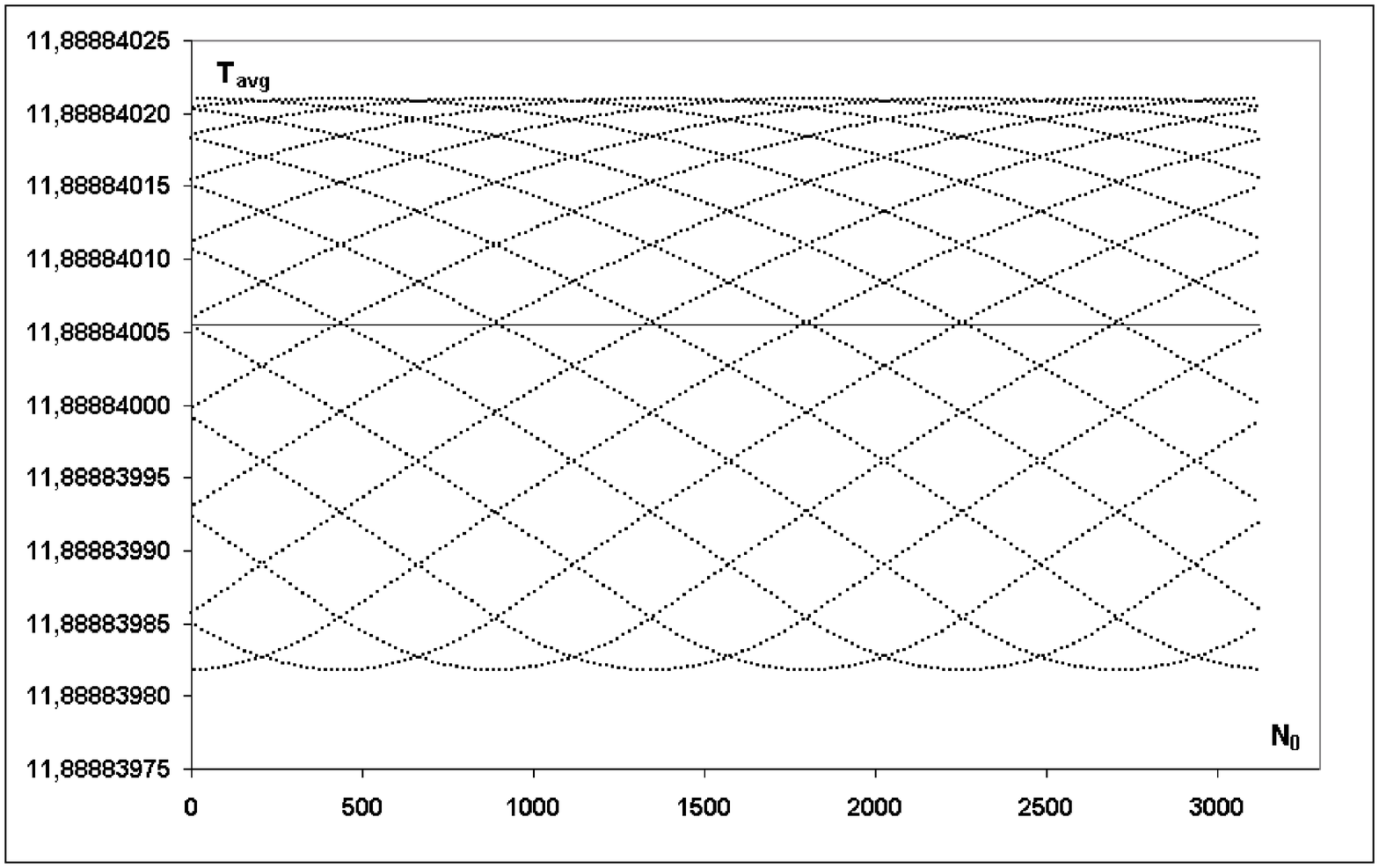} \par
\end{figure}

\begin{figure}
\caption{\small $T_{avg} (N_0, 20)$  for the Suris1 scheme (for very large $N_0$), $\ep = 0.2$, $p_0 = 1.95$,  $T_{th} = 11,65758528$, 
$T = 11,88884005$. } \label{Suris1-195-02-long} \par
\includegraphics[height=0.35\textheight]{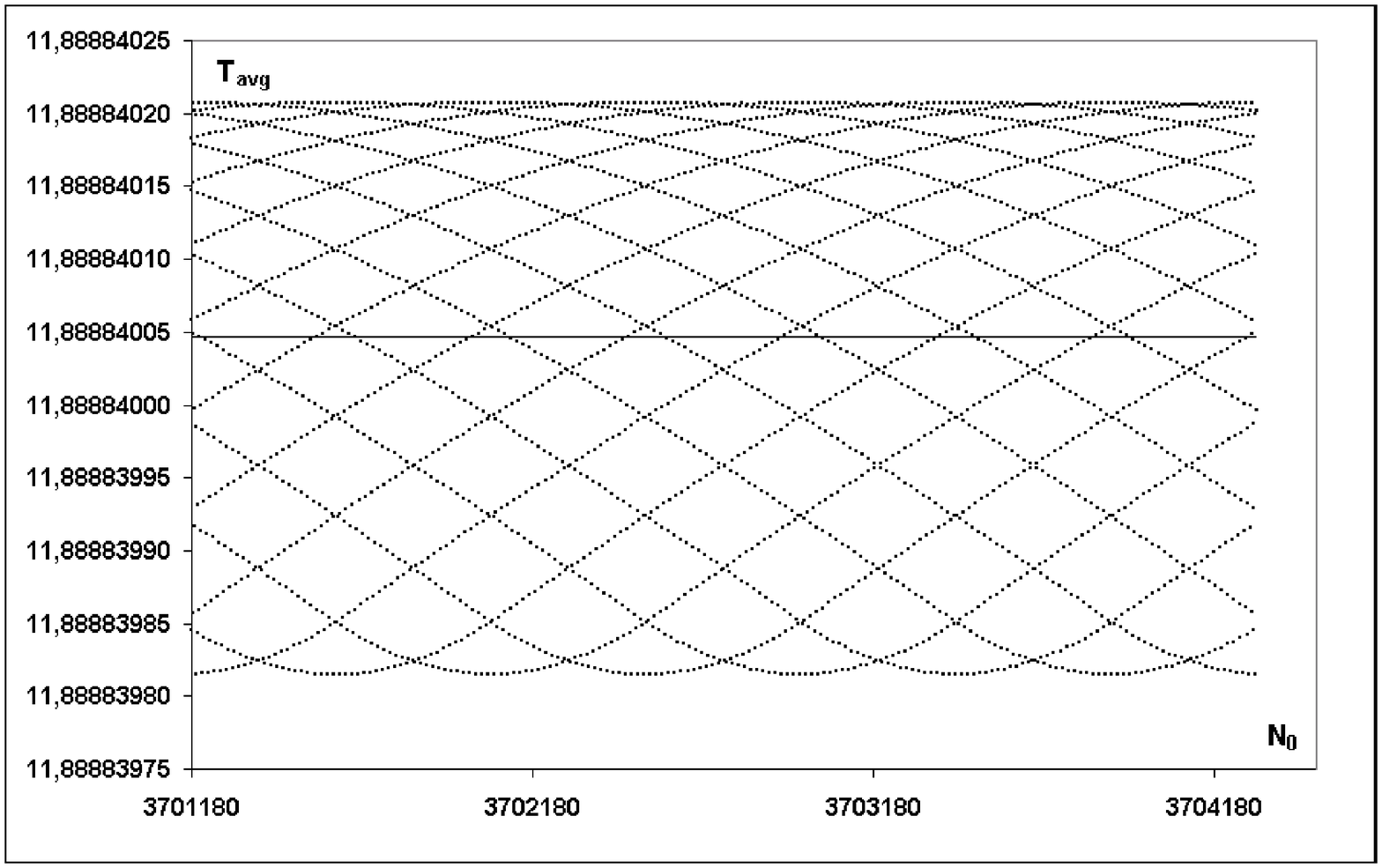} \par
\end{figure}

\begin{figure}
\caption{\small $T_N$ for the Suris1 scheme (for very large $N$), $\ep = 0.2$, $p_0 = 1.95$,  $T_{th} = 11,65758528$, 
$T = 11,88884005$. } \label{Suris1-195-02-TN} \par
\includegraphics[height=0.35\textheight]{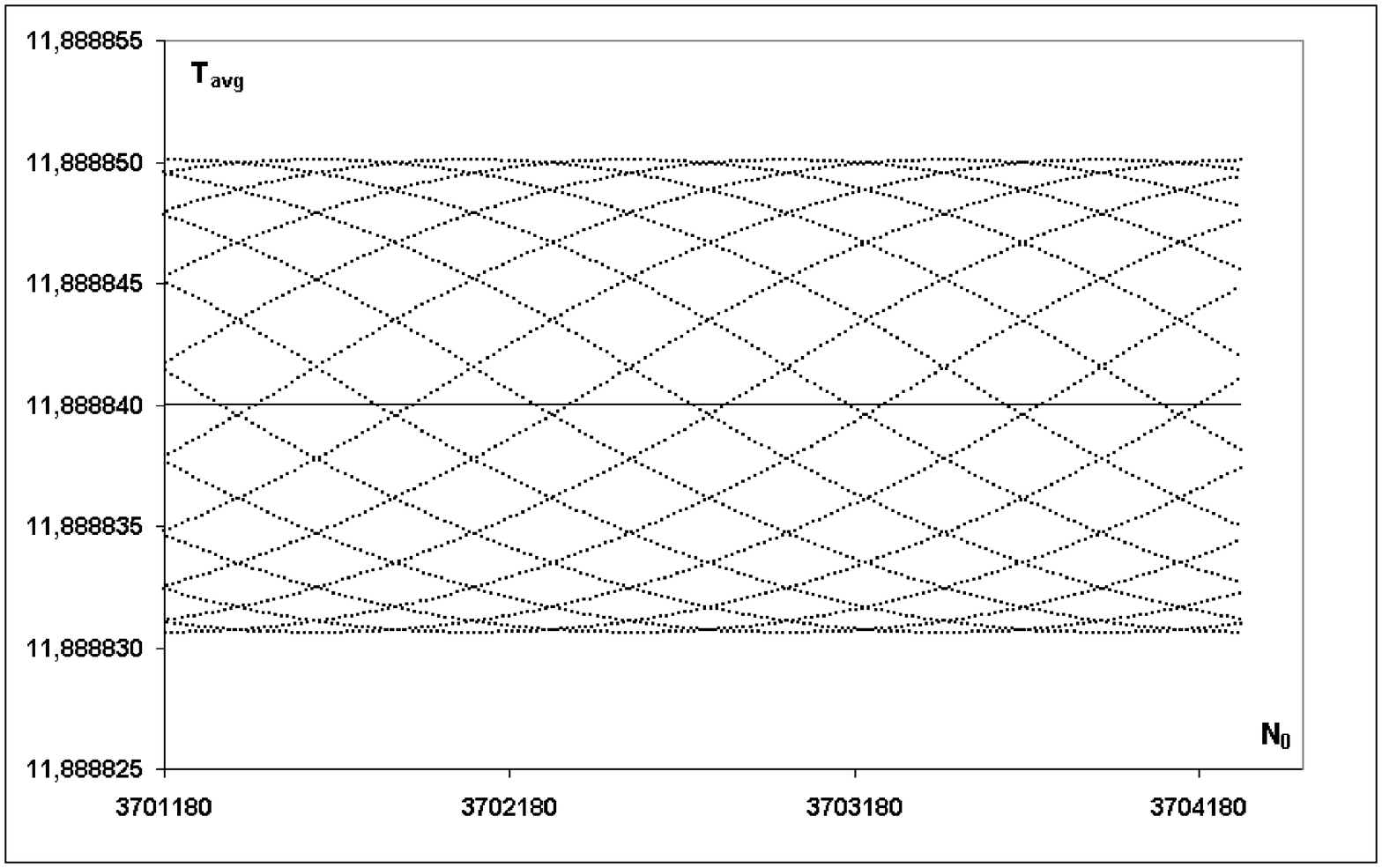} \par
\end{figure}

As a typical example we present long-time behaviour of the Suris1 scheme, see Fig.~\ref{Suris1-195-02} and 
Fig.~\ref{Suris1-195-02-long}, where we used the definition \rf{TNM} with $M=20$. 
An interesting phenomenon is associated with changing $M$. The pictures for different $M$ usually 
are very similar but the amplitude of oscillations becomes 
smaller and smaller for larger $M$ (compare Fig.~\ref{Suris1-195-02-TN}, where $M=1$, 
with Fig.~\ref{Suris1-195-02-long}, where $M=20$). 

Table~\ref{stability} shows how stable are periods of the oscillations. 
Maximal $T_N$ is defined as $\max_{J+100 < N \leqslant J +200} T_N$ for either $J=0$ or $J=1.8 \cdot 10^6$. 
Minimal values and the average are taken over the same range of values. The standard error of the average is 
about $5.7 \cdot 10^{-8}$ (the maximal error is about $10^{-7}$). Therefore, the average period is practically constant for all studied 
discretizations.  The Suris1 scheme is exceptionally stable. In this case any variations of the period are well 
within the error limits and we did not observe any dependence of $T_{avg} (N, M)$ on $N$. 
Taking into account the observed stability of the period, throughout this paper 
we identify the average period with $T \equiv T_{avg} (0, 20)$.

\begin{figure}
\caption{\small $T_{avg} (N_0,20)$ for a 4th order Runge-Kutta scheme, $\ep = 0.2$, $p_0 = 1.95$, $T_{th} = 11,65758528$. 
 } \label{RK-195-02} \par
\includegraphics[height=0.35\textheight]{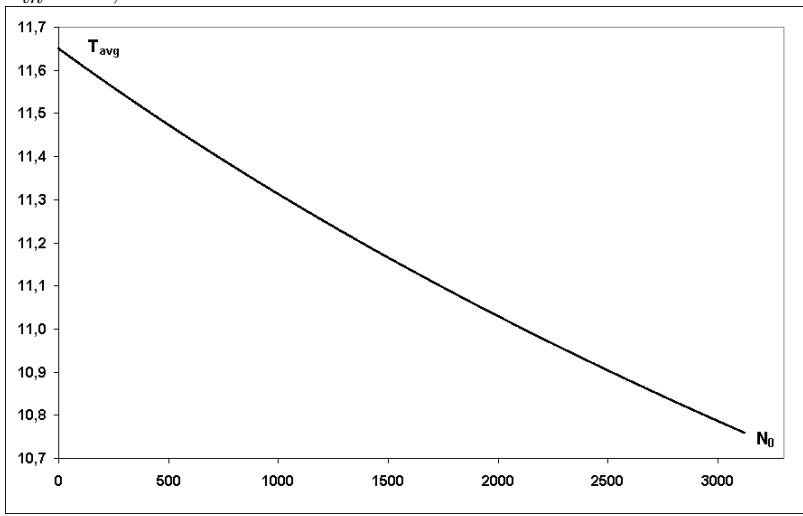} \par
\end{figure}

The observed stability of the period (for symplectic and integrable discretizations) 
is in sharp contrast with the results given by  standard 
(non-symplectic and non-integrable) numerical methods. For instance, the most popular (explicit) 4th order Runge-Kutta scheme 
yields the period noticeably decreasing in time (see Fig.~\ref{RK-195-02}). For small $N_0$ we get reasonably good estimation 
of the period (interpolating the discrete curve we get $T = 11,64602$ for $N_0=0$, which is quite close to the theoretical 
value $T_{th} = 11,65758528$. From among our discretizations only both gradient schemes produce 
comparable (even a little bit better) results, namely the discrete gradient scheme yields $T= 11.64698$. 
However, for larger $N_0$ the Runge-Kutta method yields 
worse and worse estimation of the period (in fact this is an 
exponential decrease, although very slow) while both gradient methods remain stable for very long time, 
compare Table~\ref{stability}. In this particular case ($p_0 = 1.95$, $\ep = 0.2$) 
the error produced by the Runge-Kutta 
method becomes greater than the errors of all methods considered in this paper beginning from 
$N_0 \approx 2000$.

\begin{figure}
\caption{\small Relative amplitude of the period oscillations ($\tau$) for $p_0 = 1,8$. Black circles: symmetric projection, 
pluses: standard projection, white diamonds: discrete gradient,  black squares: Suris1, stars: modified discrete gradient, black triangles: 
Suris2, black circles: leap-frog, dashed line: implicit midpoint.  } \label{Tosc-18} \par
\includegraphics[height=0.4\textheight]{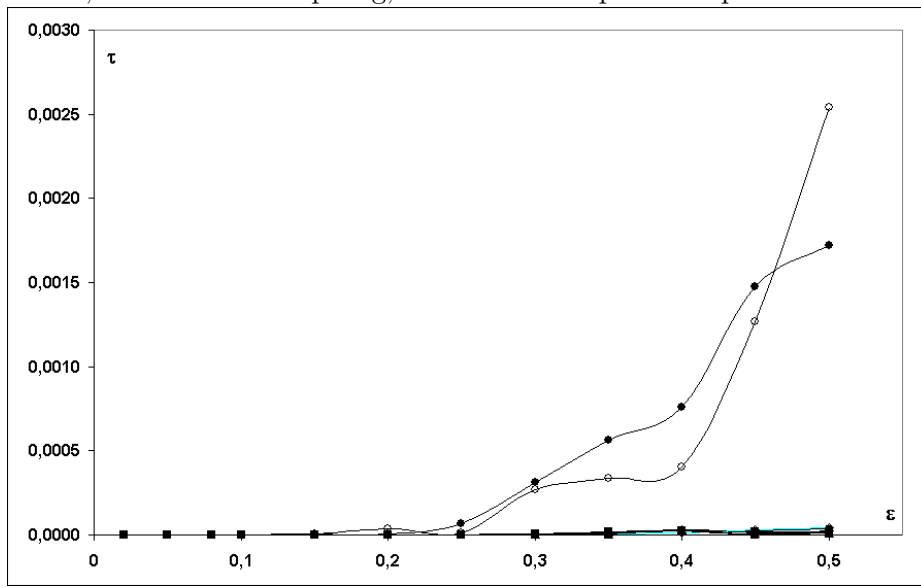} \par
\end{figure}

Numerical experiments show that the oscillations of the period and the amplitude are very small. 
For $\ep \rightarrow 0$ we have $\tau (\ep, p_0) \rightarrow 0$, up to the 
round-off error. 
The largest values of $\tau (\ep, p_0)$, obtained for both projection methods (for large $\ep$
and small $p_0$), are of order 0.2. All other discretizations
yield oscillations smaller by one or two orders of magnitude
(even for large $\ep$). A typical picture is given at Fig.~\ref{Tosc-18} representing $\tau (\ep, p_0)$ for 
$p_0 = 1.8$.

\section{Why the period and the amplitude oscillate in a very regular way?}

In a large range of parameters the oscillations $\tau_N$ are very regular and their amplitude
is greater than numerical errors by several orders of magnitude. 
This phenomenon turns out to be caused mainly by systematic numerical by-effects. 

\begin{figure}
\caption{\small $T_N$ for the leap-frog scheme, $\ep = 0.05$, $p_0 = 1.8$, 
$T = 9,1254145545$. } \label{T-leapfrog-005-18} \par
\includegraphics[height=0.45\textheight]{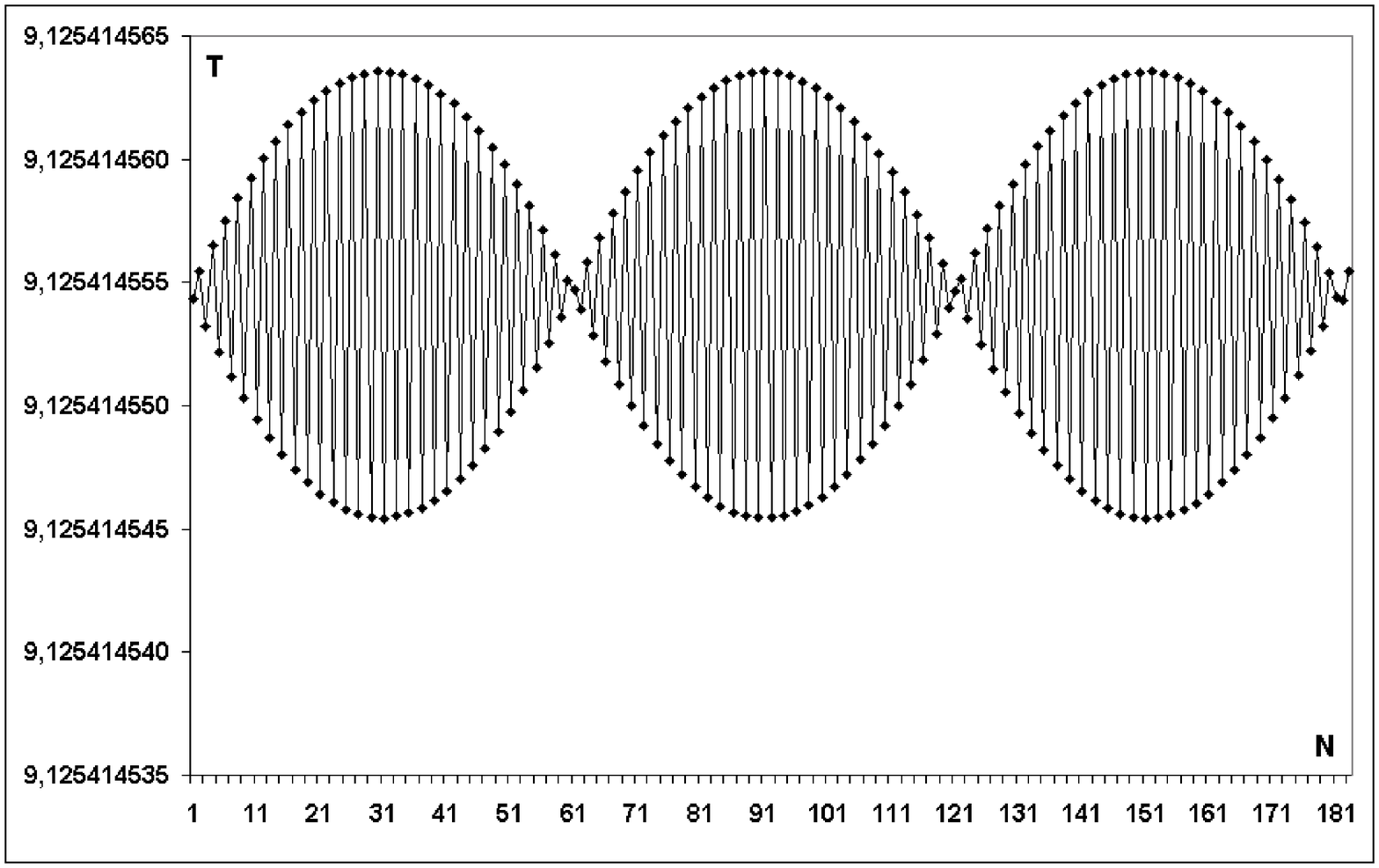} \par
\end{figure}

\begin{figure}
\caption{\small $A_N$ for the leap-frog scheme, $\ep = 0.05$, $p_0 = 1.8$, $T = 9,1254145545$. } \label{A-leapfrog-005-18} \par
\includegraphics[height=0.45\textheight]{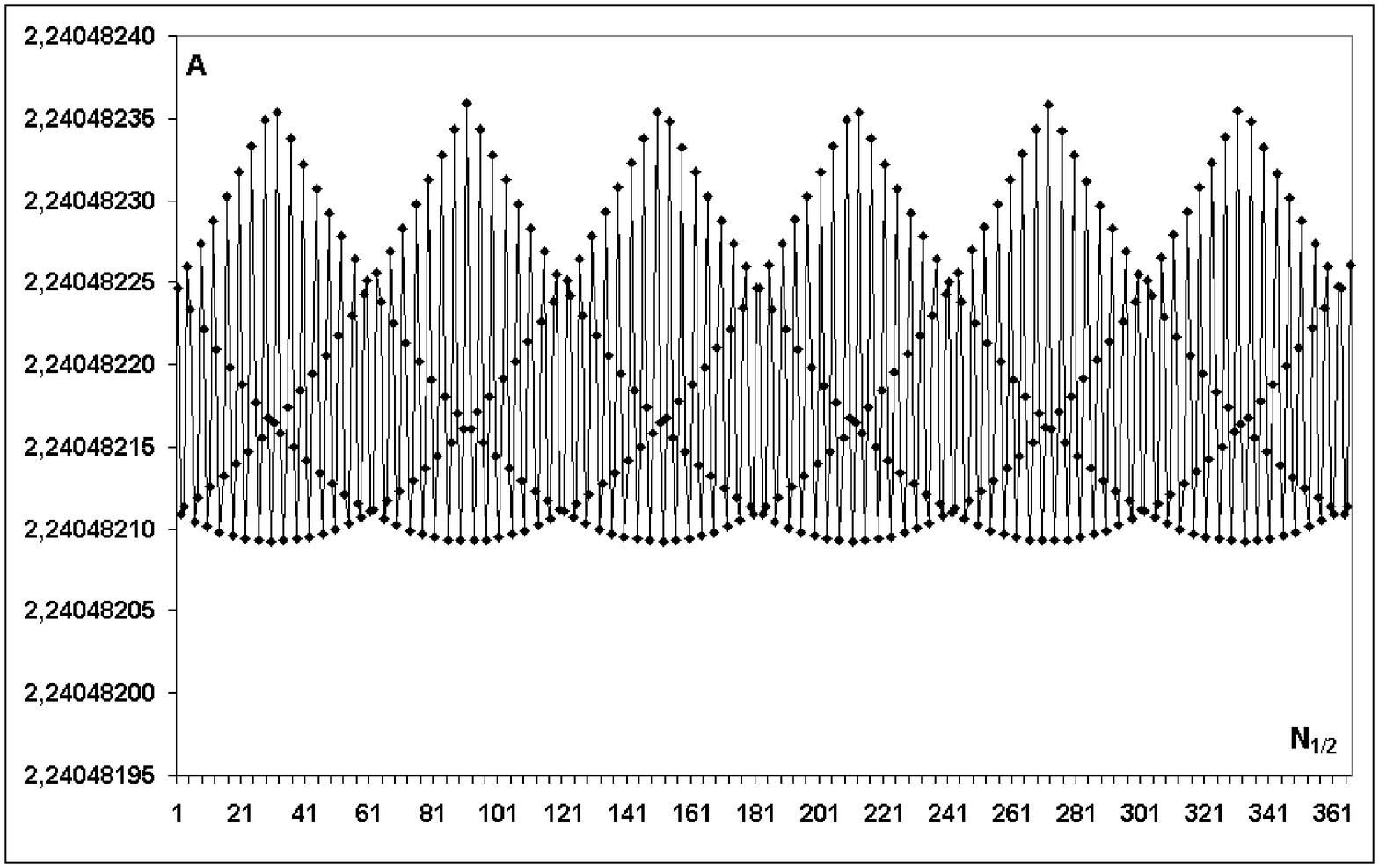} \par
\end{figure}

\begin{figure}
\caption{\small $T_N$ for the leap-frog scheme, $\ep = 0.1$, $p_0 = 0.05$, 
$T = 6,2815504224$. } \label{T-leapfrog-01-005} \par
\includegraphics[height=0.45\textheight]{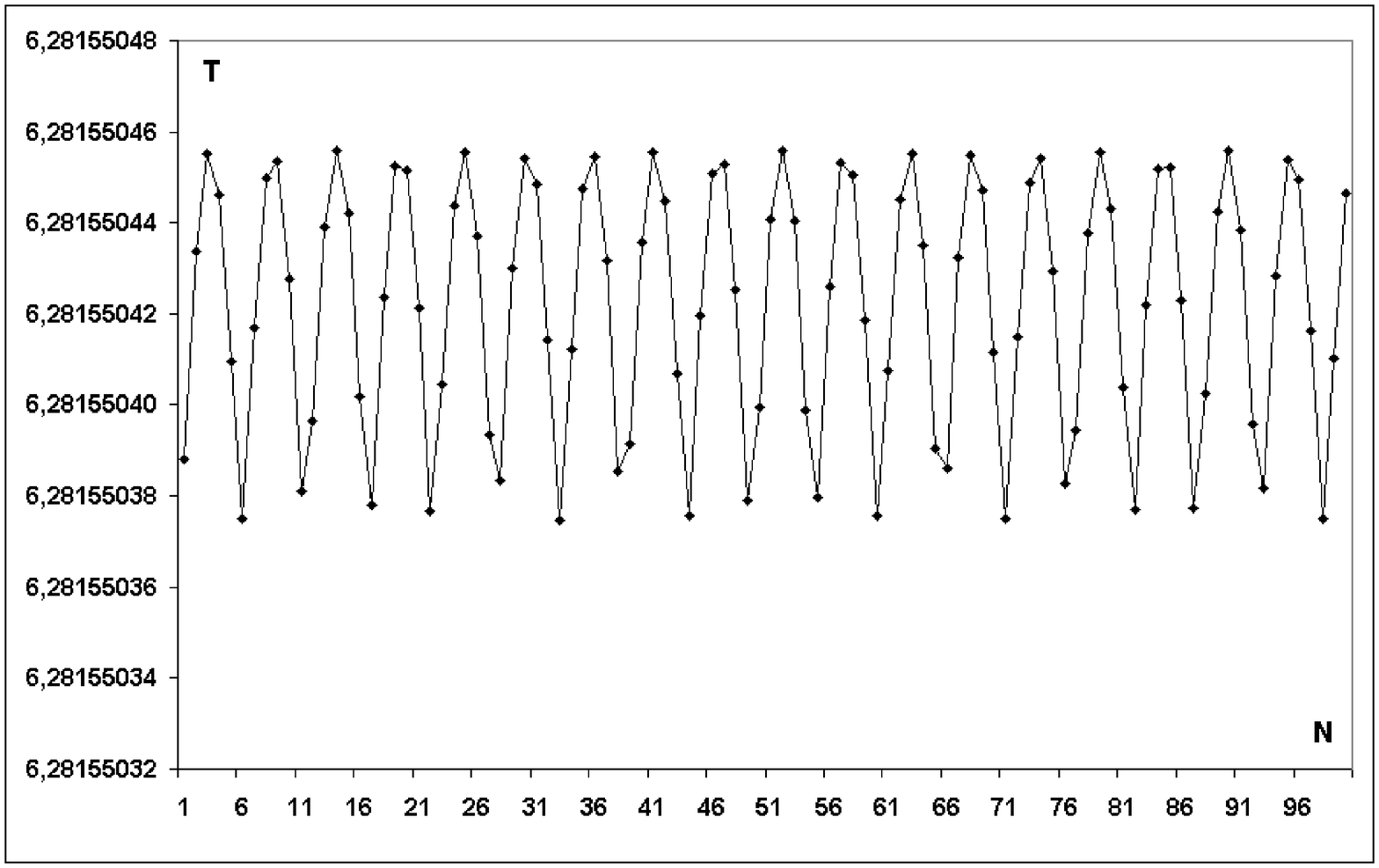} \par
\end{figure}

\begin{figure}
\caption{\small $A_N$ for the leap-frog scheme, $\ep = 0.1$, $p_0 = 0.05$, 
$T = 6,2815504224$. } \label{A-leapfrog-01-005} \par
\includegraphics[height=0.45\textheight]{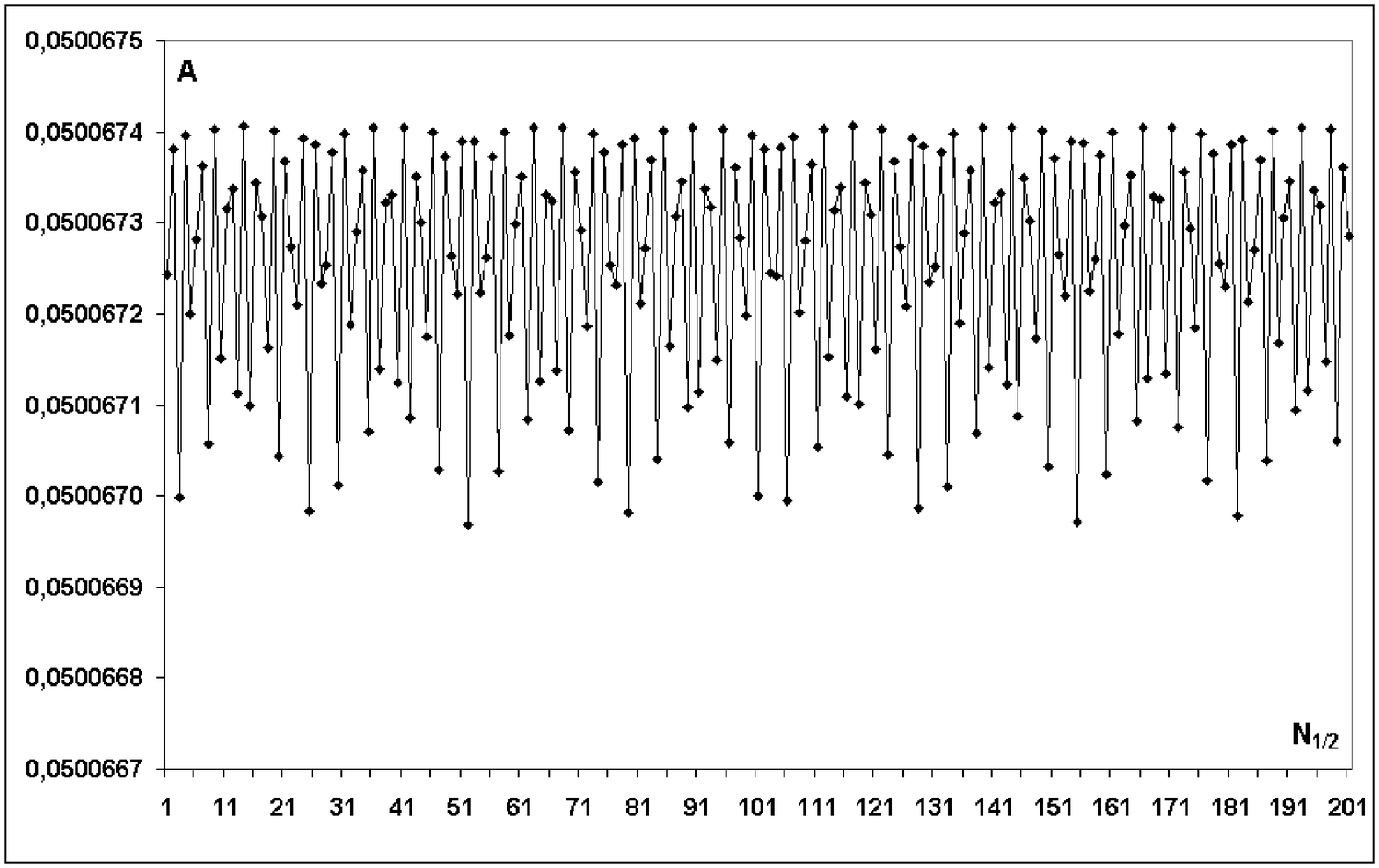} \par
\end{figure}

\begin{figure}
\caption{\small $T_N$ for the Suris1 scheme, $\ep = 0.1$, $p_0 = 0.05$, 
$T = 6,297237955$.}  
\label{T-Suris1-01-005} \par
\includegraphics[height=0.45\textheight]{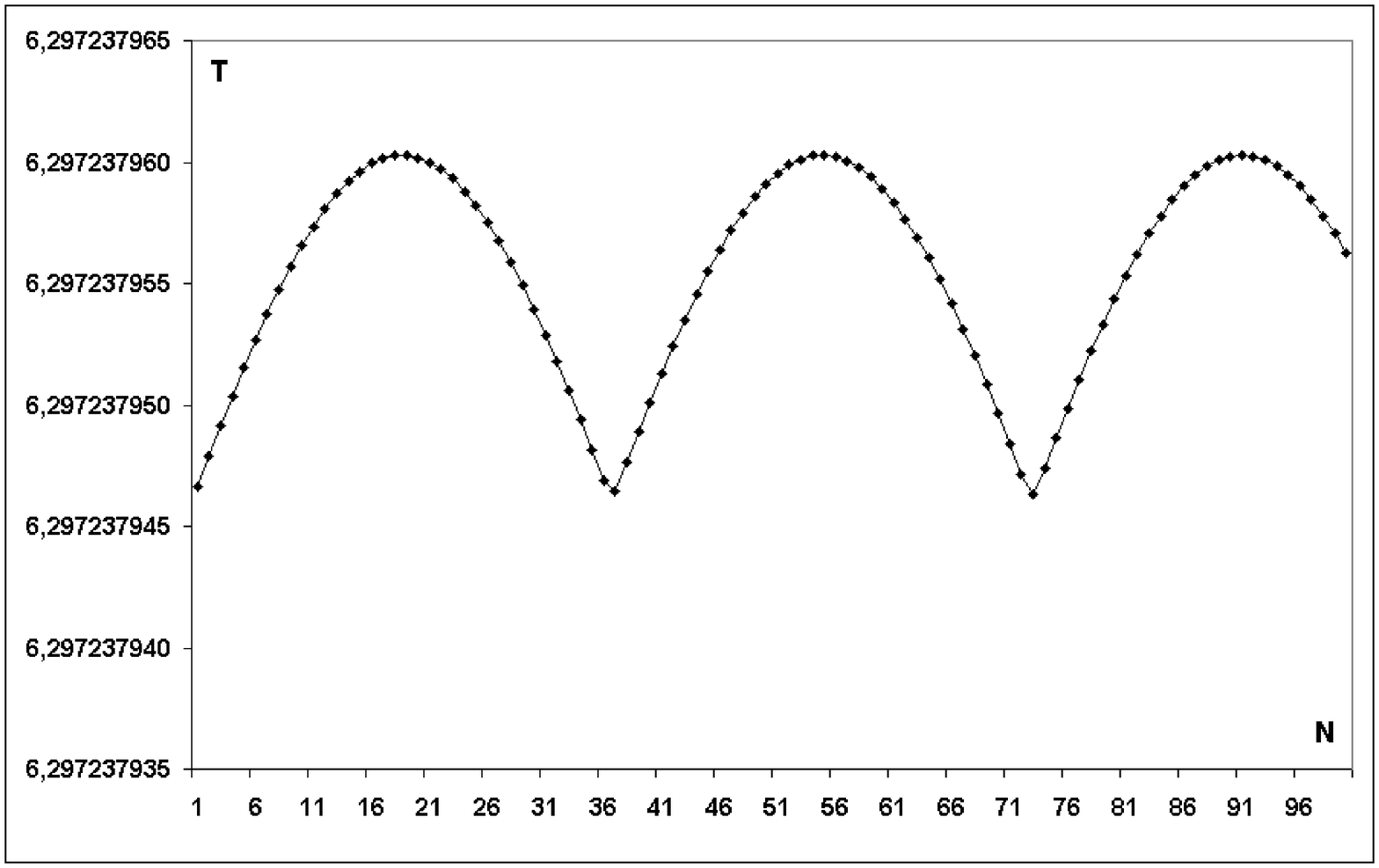} \par
\end{figure}

\begin{figure}
\caption{\small $A_N$ for the Suris1 scheme, $\ep = 0.1$, $p_0 = 0.05$, $T=6,297237955$.}  
\label{A-Suris1-01-005} \par
\includegraphics[height=0.45\textheight]{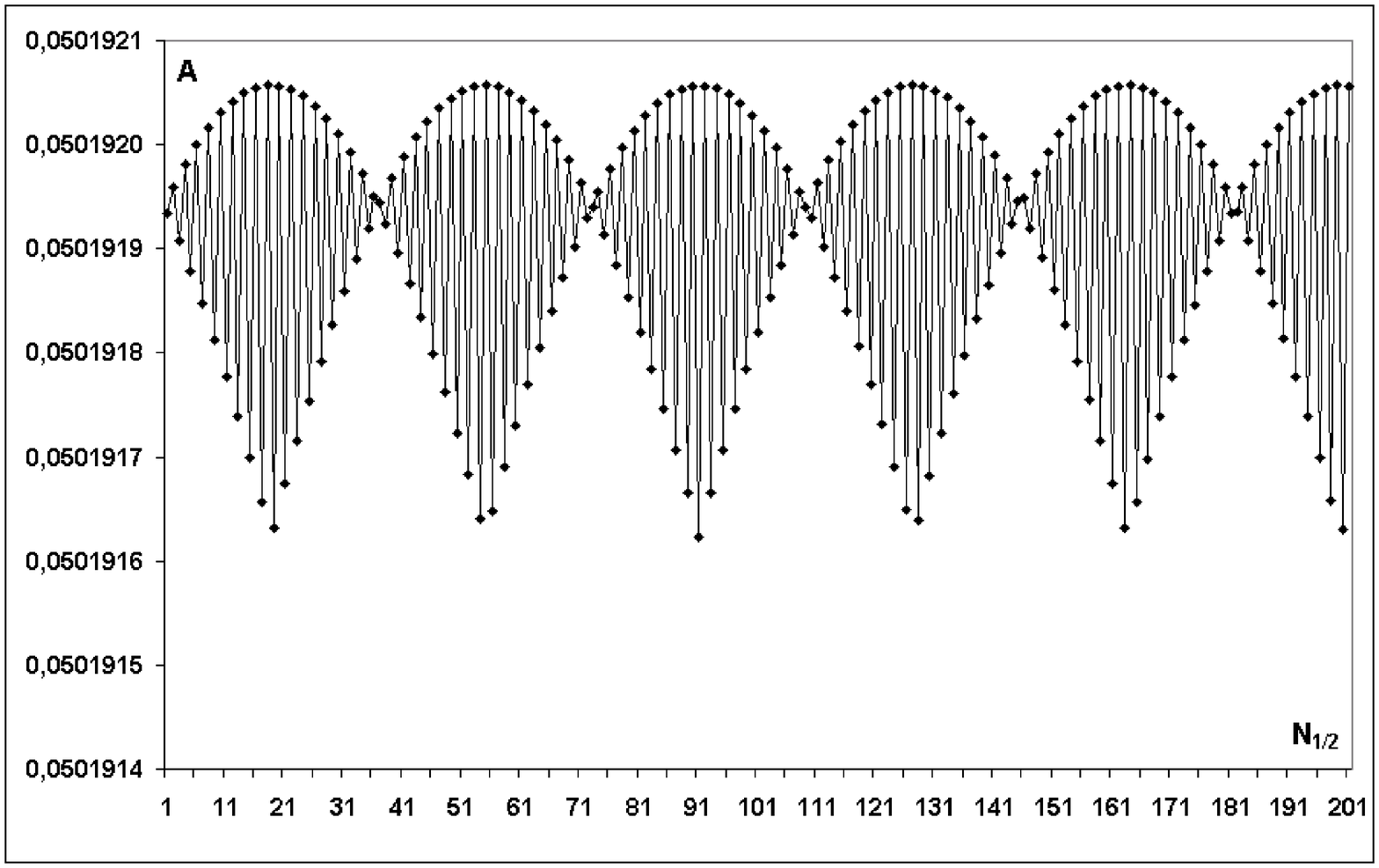} \par
\end{figure}

Our explanation is associated with the above
procedure of estimating zeros.
In general, the period $T \equiv T_{avg}$ and $\ep$ are incommensurable.
Therefore the relative position of $z_N$ between $\ph_{m}$ and $\ph_{m+1}$
depends on $N$. 
We conjecture that the periodic phenomena one observes at Fig.~\ref{T-leapfrog-005-18}, Fig.~\ref{A-leapfrog-005-18}, 
Fig.~\ref{T-leapfrog-01-005}, Fig.~\ref{A-leapfrog-01-005}, Fig.~\ref{T-Suris1-01-005} and Fig.~\ref{A-Suris1-01-005} 
are associated with the properties of the real number $T/\ep$, namely, with the approximation of 
$T/\ep$ and $T/(2\ep)$ by rational numbers. 

We begin with a simple definitions. Given  $T, \ep \in \R$ ($T > \ep > 0$) and $K \in \N$ we define:
\be  \label{munu} 
 \mu_K := \frac{K T}{\ep} - M_K  \ , \quad \nu_K := \frac{K T}{2 \ep} - L_K \ ,   
\ee
such that $- 0.5 < \mu_K \leqslant 0.5$, $- 0.5 < \nu_K \leqslant 0.5$ and $M_K, L_K \in \N$.
In other words, for a given $K$ we take $M_K$ such that $M_K/K$ is the best rational approximation 
(with a given denominator $K$) of the real number $T/\ep$, and $L_K/K$ is the best rational approximation 
(with the denominator $K$) of $T/\ep$. For  given $T, \ep, K$  the formulas \rf{munu} define uniquely 
$\mu_K$, $\nu_K$, $M_K$, $L_K$.  
The following lemma can be derived directly from the above definitions.

\begin{lem} \label{lemmunu}
Suppose that $T > \ep > 0$ are  given. 
\begin{enumerate}
\item If \ $|\mu_K + \mu_J| < 0.5$, then $M_{K+J} = M_K +M_J$ and $\mu_{K+J} = \mu_K + \mu_J$.
\item If \ $|\nu_K + \nu_J| < 0.5$, then $L_{K+J} = L_K +L_J$ and $\nu_{K+J} = \nu_K + \nu_J$.
\item If \ $|\nu_K| < 0.25$, then $M_K = 2 L_K$ and $\mu_K = 2 \nu_K$.
\item If \ $K$ is even, then $M_{K/2} = L_K$ and $\mu_{K/2} = \nu_K$. 
\end{enumerate}
\end{lem}

\begin{cor}
If $\nu_K \approx 0$, then $\mu_K \approx 0$ and, for $K$ even, also 
$\mu_{K/2} \approx 0$. 
\end{cor}

If $\mu_K \approx 0$, then the configuration
of $z_N$, $\ph_m$, $\ph_{m+1}$ practically repeats after every $K$
periods.  Therefore it is natural to expect 
some periodic recurrences
with the period $K T$. In particular, $\tau_{N+K} \approx \tau_N$ for any $N$. 
 
To obtain a ``good'' approximation we usually demand at least $\mu_K < 0.01$. 
Sometimes, especially for small $K$ (e.g., $K \leqslant 5$), interesting effects can be observed also for 
larger $\mu_K$ (but, anyway, $\mu_K < 0.1$): the graph of the function $N \rightarrow T_N$ 
apparently splits into $K$ ``discrete curves'' ($T_N$ and $T_M$ belong to the same curve if $N = M \ ({\rm mod} K$)).

Similar considerations can be made for the oscillations $\alpha_N$ of the amplitude. 
In this case the period is $T/2$ and ``good'' approximations correspond to $\nu_K \approx 0$.

\begin{ex}[\rm leap-frog scheme, $\ep = 0.05$, $p_0 = 1.8$, $T \approx 9.1254146$] \label{Ex1}
We compute $T/\ep \approx 182.508291$ and 
easily check that 
$\mu_2 \approx 0.017$, $\mu_{59} \approx - 0.011$, $\mu_{61} \approx 0.0058$, $\mu_{120} \approx - 0.0051$, $\mu_{181} \approx 0.00067$.
Fig.~\ref{T-leapfrog-005-18} confirms that the characteristic "time scales" responsible for
the pattern of the oscillations are 2, 120, and 181, indeed.

The period 2 corresponds to oscillations between two
sinusoid-like curves. Namely, $T_N$ belong to the first ``sinusoid'' 
for $N$ odd, and to second ``sinusoid'' for $N$ even. Both discrete curves are periodic with the period 120. 
Actually, the whole picture seems to have the translational symmetry with the period 60. 
The difference between $T_{N+60}$ and $T_N$ is quite large (in this sense 60 is not a period, indeed), 
however $T_N$ lays between $T_{N+59}$ and $T_{N+61}$.

The next period, 181, is more dificult to be noticed and
 corresponds to more subtle effects, like the configuration of  points near  intersections of both
"sinusoids" which approximately repeats every three 
"sinusoid"-half-periods.

Similarly, we compute  
$\nu_4 \approx 0.017$, $\nu_{59} \approx - 0.0054$, $\nu_{181} \approx 0.00034$ and $\nu_{240} \approx -0.0051$. 
On Fig.~\ref{A-leapfrog-005-18} we recognize four discrete curves, periodic with the period 240. 
The whole picture has the period 60 but looking closely on some details (e.g., at peaks or at intersections) 
we can also notice another periodicity with the period 181.

Finally, we point out that all equalities suggested by Lemma~\ref{lemmunu} hold (e.g., $\mu_{61} = \mu_2 + \mu_{59}$, 
$\nu_{240} = \nu_{59} + \nu_{181}$, $\mu_{59} = 2 \nu_{59}$, $\mu_4 = \nu_2$ etc.). 
\end{ex}

\begin{ex}[\rm leap-frog scheme, $\ep = 0.1$, $p_0 = 0.05$, $T \approx 6.28155042$] \label{Ex2}
$T/\ep \approx 62.815504$ and we check that 
$\mu_5 = 0.078$, $\mu_{11} = - 0.029$, $\mu_{27} = 0.019$, $\mu_{38} = - 0.011$, $\mu_{65} = 0.0078$, 
$\mu_{103} = - 0.0031$.    
Fig.~\ref{T-leapfrog-01-005} does not look so regularly as Fig.~\ref{T-leapfrog-005-18}. 
Note that $\mu_K$ 
are now relatively large, the first $\mu_K$ smaller that $0.01$ has the index $K=65$ and the next one is $K=103$. 
However, a 
closer inspection reveals similar features in both figures. We have five sinusoid-like curves (periodic with the period 
65). The distance between them is 13 but the difference between $T_{N+13}$ and $T_N$ is large. Note that the period 
$103 \approx 8 \times 13$, so points of only every eighth ``sinusoid'' practically coincide.  

The other periods ($K = 11, 27, 38$) can be derived from $103$ and $65$, namely: $38 = 103-65$, $27=65-38$, $11=38-27$. 
They can be noticed on Fig.~\ref{T-leapfrog-01-005} as well. For instance, the lowest points ($T_N$ between 6.28155037 and 6.28155038) 
have $N = 6, 17, 22, 33, 44, 49, 60, 71, 82, 87, 98$, the distances between them are 
given by $\Delta N = 11, 5, 11, 11, 5, 11, 11, 11, 5, 11$ 
(note that $11 + 11 + 5 = 27$). 

To explain regularities on Fig.~\ref{A-leapfrog-01-005} we compute 
$\nu_5 = 0.039$, $\nu_{22} = - 0.029$, $\nu_{27} = 0.0093$, $\nu_{49} = - 0.020$, 
$\nu_{76} = - 0.011$, $\nu_{103} = - 0.0015$, $\nu_{130} = 0.0078$ and also $\nu_{645} = 0.00010$. 
In this case the structure is also quite complicated because we have
several candidates for periods. Some of them admit
a clear interpretation.
Joining every fifth point we get five sinusoidal curves
 with the period 130.
Thus the distance between neighbouring ``sinusoids'' is 26 which is very close to the period 27. 
The subsequent minima are at $N = 3, 25, 52, 79, 106, 128, 155, 182, 209$, therefore $\Delta N = 22, 27, 27, 27, 22, 
27, 27, 27$ (note that $|\nu_{22}|$ is also relatively small). Looking at configurations of points near every minimum we can notice 
a distinct periodicity with the period 103. 
\end{ex}

\begin{ex}[\rm Suris1 scheme, $\ep=0.1$, $p_0=0.05$, $T \approx 6.29723795$]  \label{Ex3}

In this case the structure of Fig.~\ref{T-Suris1-01-005} is extremaly simple (a single discrete curve). 
It can be explained by the non-existence of any ``small'' periods. The smallest one, 
distinctly seen at Fig.~\ref{T-Suris1-01-005}, is 36. Namely, 
$\mu_{36} = 0.0057$, $\mu_{145} = 0.0050$, $\mu_{181} = 0.00069$. The period 
181  is even more exact than the period 36 ($\mu_{181}$ is much smaller than $\mu_{36}$). Therefore 
after every five basic periods ($181 \approx 5 \times 36$) the periodicity improves.  

Fig.~\ref{A-Suris1-01-005} consists of two intersecting discrete curves (periodic with the period 72), 
because $\nu_2 = - 0.028$ is relatively small and $\nu_{72} = 0.0057$. Actually the important point is that 
$\nu_3 = 0.46$ is 
much greater than $|\nu_2|$. Note that $\mu_2 = - 0.055$ is also not very large but 
 $\mu_3 = - 0.083$ is of the same order. The whole structure has the period 36 but (similarly as in Example~\ref{Ex1}) 
the difference between $A_{N+36}$ and $A_N$ is quite large, $A_N$ is close to $A_{N+35}$ and $A_{N+37}$ ($\nu_{35} = 0.017$, 
$\nu_{37} = - 0.011$). Moreover, we have the period 181, quite accurate ($\nu_{181} = 0.00034$). This periodicity 
can be noticed by looking at the minima or at points where the discrete curves ``intersect''.  

\end{ex}

Similar remarks concern the case presented at Fig.~\ref{Suris1-195-02}, Fig.~\ref{Suris1-195-02-long}, 
Fig.~\ref{Suris1-195-02-TN}, where 
$T \approx 11,88884005$ and 
$\mu_9 = - 0.0044$, $\mu_{448} = 0.0035$, $\mu_{457} = - 0.00093$. 
The patttern on any of these figures consists of nine discrete curves and is periodic with the period 
close to 457.

The behaviour described on the above examples is typical and similar periodic phenomena can be observed 
for other discretizations and for other choices of parameters except very small values of $\ep$ 
(e.g., $\ep \leqslant 0.01$) when periodic oscillations are comparable or smaller than the round-off error 
(then the oscillations become chaotic with a very small amplitude).

\section{Numerical estimates of the amplitude and the period}

All discretizations considered in this paper are characterized by very good stability 
of their trajectories. Therefore, such quantities as average period and  
(in the case of oscillating motions) average amplitude are well defined for 
every discretization (provided that $\ep$ is not too large, it is sufficient to assume 
$\ep \leqslant 0.5$).  

\subsection{Average amplitude}

Relative errors for the average amplitude are presented in Table~\ref{error-amp} 
(for $\ep = 0.02$ and $\ep = 0.5$). They were computed as differences between the 
numerical results and exact amplitudes given in terms of elliptic functions. 
One can immediately see that in any case the best results are given 
by both gradient schemes 
(and the worst ones are given by Suris1 and Suris2 schemes). 
The relative error of the leap-frog and Suris' methods practically 
does not depend on $p_0$. 
The accuracy of gradient methods 
increases for larger $p_0$, both for $\ep=0.02$ and $\ep=0.5$. 
For small $\ep$ (e.g., $\ep = 0.02$) also projection schemes yield very small 
errors, 
like $10^{-8}$ or $10^{-9}$ (similar as gradient 
methods). However, for some $p_0$ their accuracy is very high (e.g, for $p_0 =1.6$) while 
for some other $p_0$ -- relatively worse (e.g., for $p_0 = 0.8$). 

The implicit midpoint rule is comparable 
to gradient methods but only for small $p_0$ (e.g., 
$p_0 < 0.1$). The leap-frog method, both Suris' discretization 
and (for $p_0 > 1.6$) the 
implicit midpoint rule yield much larger errors (by 4 orders of magnitude). 

For greater $\ep$ (e.g., $\ep = 0.5$) the differences between the studied methods 
are much smaller (they differ at most by 2 orders of magnitude). 
Gradient methods are most accurate. 
The implicit midpoint rule has similar accuracy for $p_0 < 1.2$ while 
projection methods are not much worse for $p_0 > 1.8$. 
Leap-frog method and both Suris' methods have larger relative errors for any $p_0$. 
We point out, however, that even those ``large'' errors are not so bad (only several percent) 
with the exception of $p_0$ approaching $2$ (when these discretizations fail to 
reproduce properly even the qualitative behaviour).

\begin{figure}
\caption{\small $A_{avg} \equiv A_{avg} (0,50)$ as a function of $\ep$ for  $p_0 = 1,8$ ($A_{th} = 2,239539$). 
White triangles: leap-frog, black triangles: implicit midpoint, white squares: Suris1, black squares: Suris2, 
black diamonds: modified discrete gradient (discrete gradient method yields practically the same values), 
black circles; symmetric projection, white circles: standard projection (usually covered by black circles). } 
\label{Aavg-18} \par
\includegraphics[height=0.35\textheight]{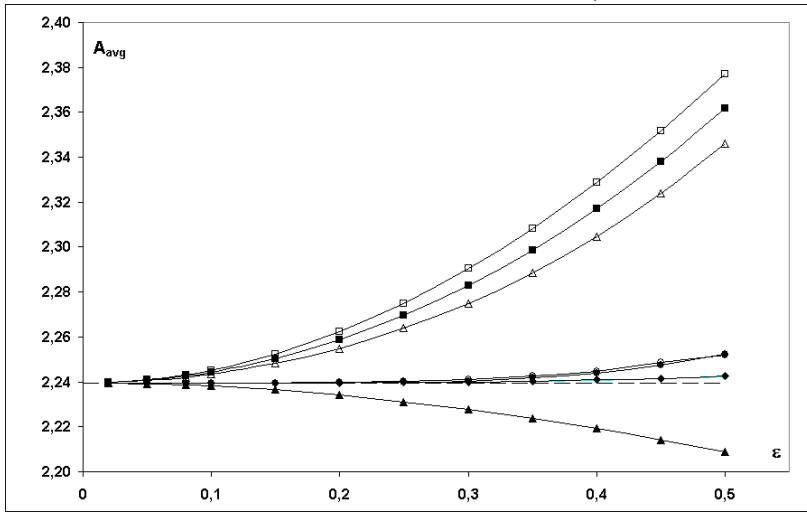} \par
\end{figure}

Fig.~\ref{Aavg-18} illustrates the dependence of the average amplitude on $\ep$ for 
$p_0 = 1.8$. Gradient methods and (especially for $\ep < 0.3$) projection 
methods are most accurate.

\subsection{Average period}

\begin{figure}
\caption{\small Modified discrete gradient method. Relative error as a function of $\ep$ for 
$p_0 = 0,02$, $T_{th} = 6,283342395$, $T (\ep) = T_{avg} (0,30)$.  } 
 \label{delta-error}  \par
\includegraphics[height=0.4\textheight]{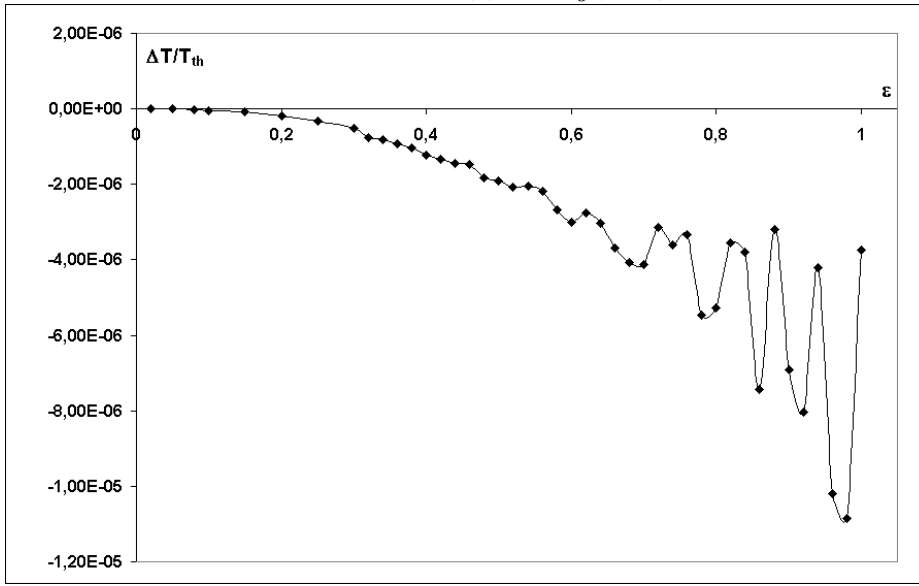} \par
\end{figure}

Relative errors for the average period are presented in Table~\ref{error-per} 
and also in Table~\ref{error-sep} (in both cases for $\ep = 0.02$ and $\ep = 0.5$). 
For $p_0 < 0.5$ all discretizations except the modified discrete gradient method 
have similar 
relative errors (Suris1 scheme is the worst among them).  
The modified discrete gradient methods is much better 
(for $p_0 \approx 0$ its error is smaller by 4 orders of magnitude, at least), 
compare Fig.~\ref{Tavg-01} ($p_0 = 0.1$) and Fig.~\ref{delta-error} ($p_0 = 0.02$). 

\begin{figure}
\caption{\small Relative error as a function of $\ep$ for $p_0 = 1.21$ (``resonance value'' 
for the leap-frog scheme), $T_{th} = 7,01866131087$, $T (\ep) = {\bar T}_{avg} (0,100,200)$. 
Black triangles: implicit midpoint, white triangles; leap-frog, 
black diamonds:  modified 
discrete gradient. } 
 \label{rezonans}  \par
\includegraphics[height=0.4\textheight]{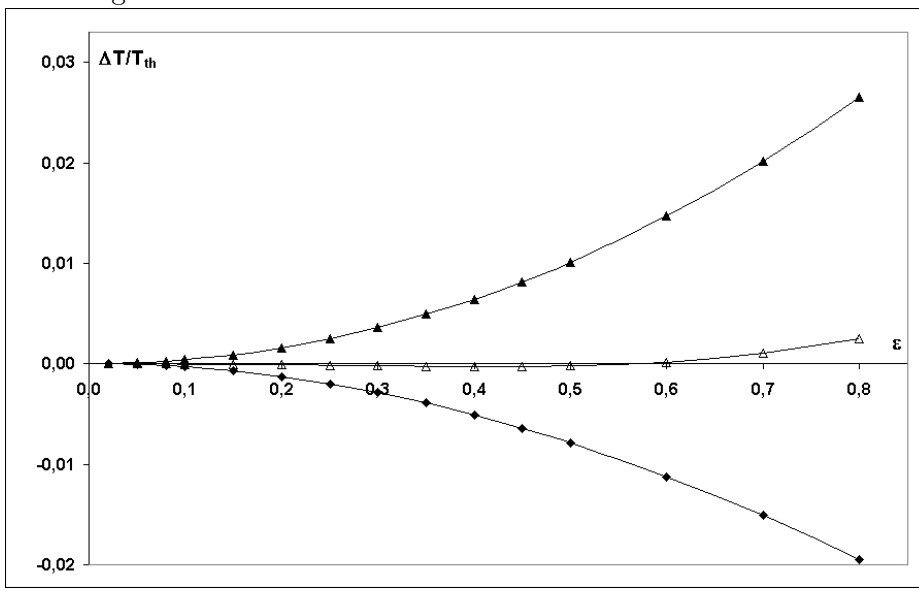} \par
\end{figure}

Then, with increasing $p_0$, all discretizations become to have similar accuracy  
with two very interesting exceptions: leap-frog and implicit midpoint schemes 
have a kind of ``resonance values'' for which their accuracy is much better than the accuracy 
of all other method. Fig.~\ref{rezonans}  shows how accurate is the leap-frog scheme for 
$p_0 = 1.21$ and for practically any $\ep$. 
There are shown also next two discretizations: implicit midpoint and modified 
discrete gradient, much worse (for this value of $p_0$) than leap-frog  
(other discretizations are even 
less accurate). 
Implicit midpoint scheme has an analogical ``resonance value'', namely  
$p_0 \approx 1.6$. It is worthwhile to point out that, surprisingly,  
projections applied to the leap frog method have strong negative effect 
on the accuracy of the average period for $0.8 < p_0 < 1.8$, 
especially for larger $\ep$ (e.g., $\ep = 0.5$). 

If $p_0$ approaches $2$, then both gradient methods become more accurate than other methods 
(only for small $\ep$ the projection methods are better). For 
$p_0$ very close to this limiting value the accuracy of all methods decreases rapidly, 
and the leap-frog method and both Suris' methods produce rotating motions instead of 
oscillations, see Table~\ref{error-sep}. The closest neighbourhood of the separatrix 
($p_0=2$) is discussed in more detail below. Here we remark only that,  for $p_0$ slightly 
greater than 2, the implicit 
midpoint method fails to reproduce rotations 
and has wrong qualitative behaviour (i.e., oscillations). 

In the case of rotating motions the relative error of the average period is very similar 
for all considered methods except the discrete gradient scheme which is better by one or 
two orders of magnitude.

\section{Interesting special cases}

In this section we briefly present several points which seem to be encouraging to  
further studies. 

\subsection{Extrapolation $\ep \rightarrow 0$}

\begin{figure}
\caption{\small $T_{avg} \equiv {\bar T}_{avg} (0,100,200)$ as a function of $\ep$ for  $p_0 = 0,1$ ($T_{th} = 6,28711782$). White squares: Suris1, 
black triangles: midpoint (Suris2 and discrete gradient methods yield practically the same results), black diamonds: 
modified discrete gradient (very close to the theoretical exact values), white triangles: leap-frog, black circles: 
symmetric projection, white circles: standard projection.  
} \label{Tavg-01} \par
\includegraphics[height=0.35\textheight]{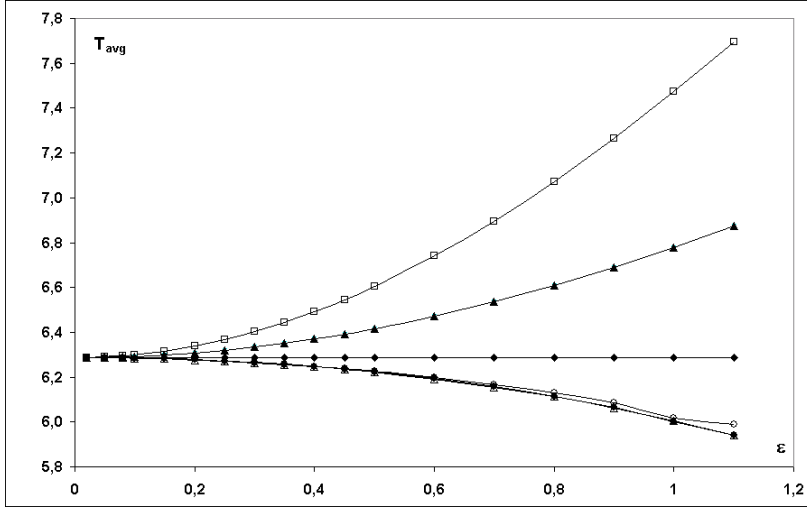} \par
\end{figure}

For all studied discretizations we expect
\be  \label{limes}
    \lim_{\ep \rightarrow 0} T (\ep, p_0) =  T_{th} (p_0) \ , 
\quad \lim_{\ep \rightarrow 0} A (\ep, p_0) =  A_{th} (p_0) 
\ee
where $T_{th} (p_0)$, $A_{th} (p_0)$ do not depend on the discretization and are equal
to theoretical values computed from the analytic formula (in terms of elliptic functions), 
compare Fig.~\ref{Tavg-01} and Fig.~\ref{Aavg-18}.  

Let us analyse quantitatively the case presented at Fig.~\ref{Tavg-01}    
(the exact period is $T_{th} \approx 6.28711783$). 
Fitting 3rd-order polynomials (very close to parabolas, in fact) 
to twelve points ($\ep = 0.01, 0.02, \ldots, 0.11, 0.12$) we get
{\small
\be \ba{l}
T = -0,03867\ep^3 + 1,310512\ep^2 - 0,0001050\ep + 6,28711875 
\quad ({\rm Suris1}) \ , \\
T = -0,00909\ep^3 + 0,524053\ep^2 - 0,0000247\ep + 6,28711805 
\quad ({\rm Suris2}) \ ,  \\ 
T = -0,00475\ep^3 - 0,260242\ep^2 - 0,0000130\ep + 6,28711794 
\quad ({\rm leap}{\rm -}{\rm  frog}) \ . \\
\ea \ee}
The last terms estimate the exact period quite well. Taking $10^{-7}$ as a unit we 
compute their absolute errors as: $9.2$, $2.2$ and $0.9$, respectively. They 
are comparable with 
the errors at $\ep = 0.001$ (given by $13.1$, $5.2$ and 
$- 2.6$, respectively). 
The errors at  $\ep=0.01$ (namely, $1307.2$, $523.3$ and $260.6$) are higher by two orders 
of magnitude. The modified discrete gradient scheme (with the $\delta$-correction) 
beats all other discretizations: 
its error at $\ep = 0.01$ is only $1.3$ (in the same units).

\subsection{The neighbourhood of the separatrix} 

The separatrix is a border between oscillating and rotational motions. 
Table~\ref{error-sep} presents the values of the period for motions near the separatrix, 
i.e., $p_0 \approx 2$. 
This is certainly the range of parameters most difficult for  
accurate numerical simulations. The gradient schemes and projection methods yield 
satisfying results, especially for small $\ep$, and are  much better than 
all other methods. For rotating motions very close to the separatrix even projection 
methods (especially the symmetric projection) become less accurate and 
only gradient methods yield relatively good 
quantitative results, see Table~\ref{error-sep}. 

The other discretizations produce wrong results (in the neighbourhood of the separatrix) 
even qualitatively. 
Namely, the leap-frog 
and both Suris' schemes begin to simulate rotating motions for $p_0 < 2$ (e.g., for $p_0 =1.99$ 
if $\ep = 0.5$, and for $p_0 = 1.99999$ if $\ep = 0.02$), while the implicit midpoint rule 
produces oscillating motions for $p_0 > 2$ (e.g., for $p_0 = 2.000001$ if $\ep = 0.02$, and 
for $p_0 = 2.001$ if $\ep = 0.5$). Even in the case of good qualitative behaviour 
these methods yield very large relative errors, 
especially for larger $\ep$  (for $\ep = 0.5$ and $|p_0 - 2| \leqslant 0.001$ leap-frog, 
implicit midpoint and both Suris' schemes yield relative errors like $30\%-70\%$ and more.

\begin{figure}
\caption{\small $\ph_n$ for $p_0=2$, $\ep = 0,2$, round-off error $\Delta=10^{-16}$.  
White circles: standard projection, black 
circles: symmetric projection, white diamonds: discrete gradient, black diamonds: 
modified discrete gradient. } 
 \label{separatrix}  \par
\includegraphics[height=0.35\textheight]{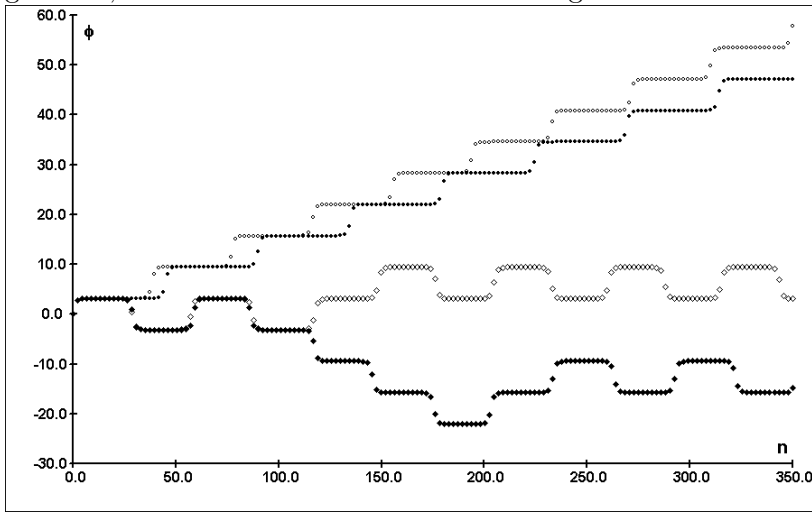} \par
\end{figure}

\begin{figure}
\caption{\small $\ph_n$ for $p_0=2$, $\ep = 0,00025$, round-off error $\Delta=10^{-18}$. 
Black squares: Suris1, black circles: symmetric projection, 
white circles: standard projection,   
black diamonds: modified discrete gradient, white diamonds: discrete gradient. 
White squares (Suris2) and white triangles (leap-frog) are almost covered by black squares.  } 
 \label{sep-all}  \par
\includegraphics[height=0.35\textheight]{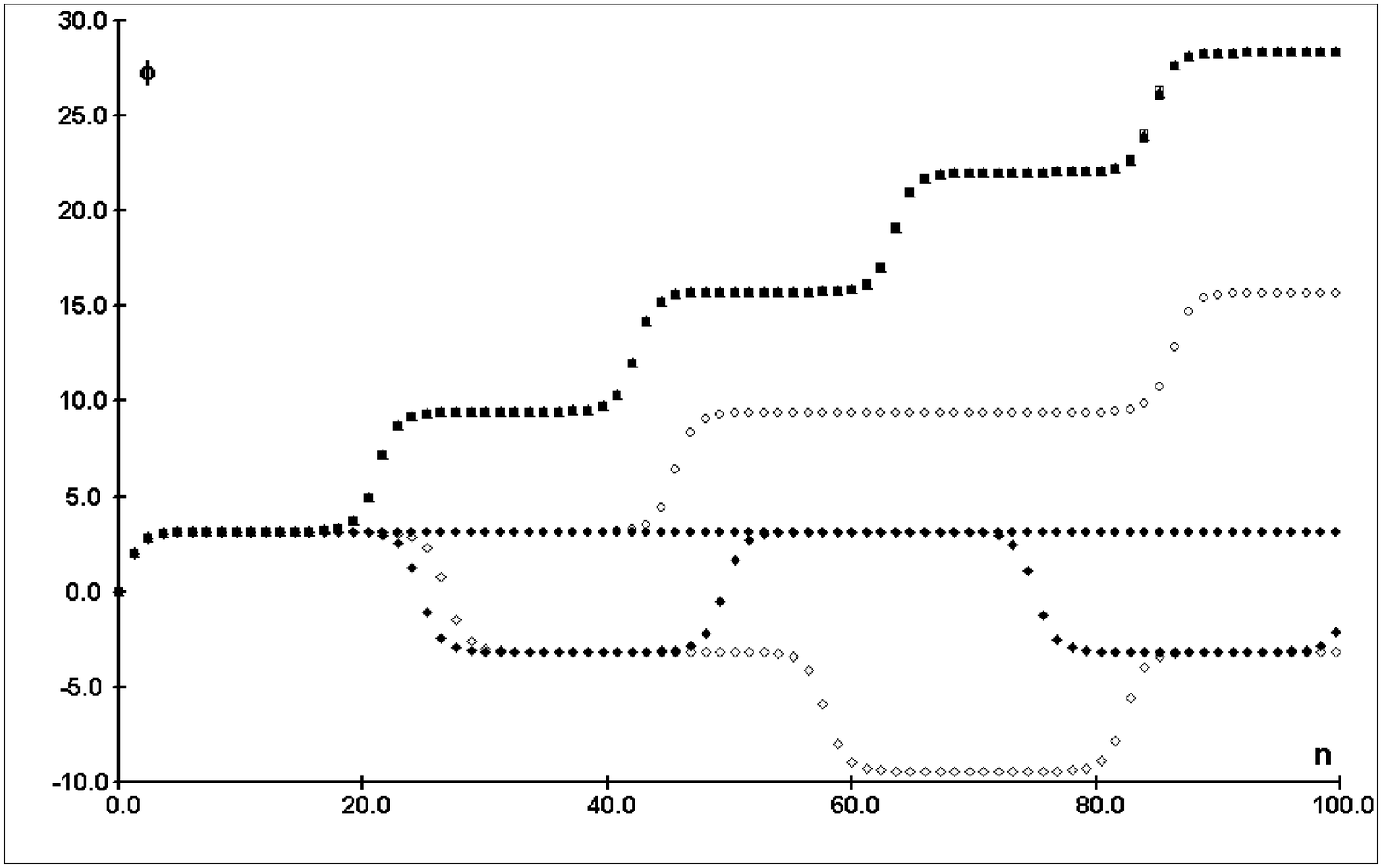} \par
\end{figure}

If $p_0 = 2$, then (in the continuous case) we have the motion along the separatrix, i.e., 
$\ph \rightarrow \pi$ for $t \rightarrow \infty$. For larger $\ep$ (e.g., $\ep = 0.2$) this behaviour is not 
reproduced by any discretization. Interesting results are given by both gradient schemes, 
see Fig.~\ref{separatrix} ($\ep = 0.2$). 
The standard gradient scheme produces oscillations, but after three periods one rotation 
is performed. The modified discrete gradient scheme gives a strange motion: first oscillations 
(two periods), then backward rotation (3 periods), forward rotation and the return to 
oscillations. This picture depends on $\ep$ and the round-off error chosen. In any case, 
for both gradient schemes,  we 
have a number of chaotic-looking switches between oscillations and rotations in both 
directions. Qualitatively this behaviour may be considered as satisfying. It reflects  
the fact that the equilibrium at $\ph = \pi$ is unstable. In the same time, the 
projective discretizations (quite good at qualitative 
description of motions near the separatrix) produce relatively slow rotational motion 
(similarly as the standard leap-frog method and both Suris schemes). However, for very small 
$\ep$ (e.g., $\ep \leqslant 0.00025$) the symmetric projection method seems to have the proper 
qualitative behaviour and is much better than other considered numerical schemes, see 
Fig.~\ref{sep-all}.

\subsection{Advantages of the new method}

The discrete gradient method with $\delta$-correction turned out to
be very efficient as far as the numerical estimation of the period (for
relatively small amplitudes) is concerned. The range of these "small"
amplitudes is quite large, up to $\ph \approx \pi/4$, which corresponds
to $p_0 < 0.8$. Thus it contains also the cases which cannot be
approximated by the linear oscillator. 
Even for $p_0 \approx 0.8$ the new method is several times better than the
best of other considered schemes, and for smaller $p_0$ it becomes better even 
by 4 orders of magnitude (e.g., for $p_0 = 0.02$ the errors of other 
discretizations are greater by the factor at least $0.5 \cdot 10^{4}$,  
see Table~\ref{error-per}).

Fig.~\ref{Tavg-01} ($p_0 = 0.1$) shows how precise is the period given 
by our new method in comparison to the period given by other numerical schemes. 
Similarly, Fig~\ref{delta-error} presents the relative error 
 for $p_0 = 0.02$ and a large range of $\ep$. 
We see that even for $\ep =1$ the relative error is only $10^{-5}$! 
For small $\ep$ the error is 
$10^{-9}$ and less. 

Our method works very well also for larger amplitudes, but for $p_0$ larger 
than $1,4$ the discrete gradient method is better, and the leap-frog scheme 
and implicit midpoint are unbeatable around their "resonance" amplitudes ($p_0 \approx 1,2$ 
and $p_0 \approx 1,6$, respectively). In the case $p_0 > 2$  the delta correction 
have negative influence on the accuracy of the gradient discretization (which is the 
best for rotating motions). However, the accuracy of the modified 
discrete gradient method is on the same level as the accuracy of all 
other considered methods. 

\begin{figure}
\caption{\small  $\ph_n$ for $p_0 = 2,000001$, $\ep = 0,1$. White triangles: leap-frog, 
black diamonds: modified discrete gradient, 
white diamonds: implicit midpoint. The period of the exact solution 
(continuous line): $T_{th} = 16,58809538$, the average period given by the modified 
discrete gradient scheme: $T = 16,56380722$. } 
 \label{2E-06}  \par
\includegraphics[height=0.45\textheight]{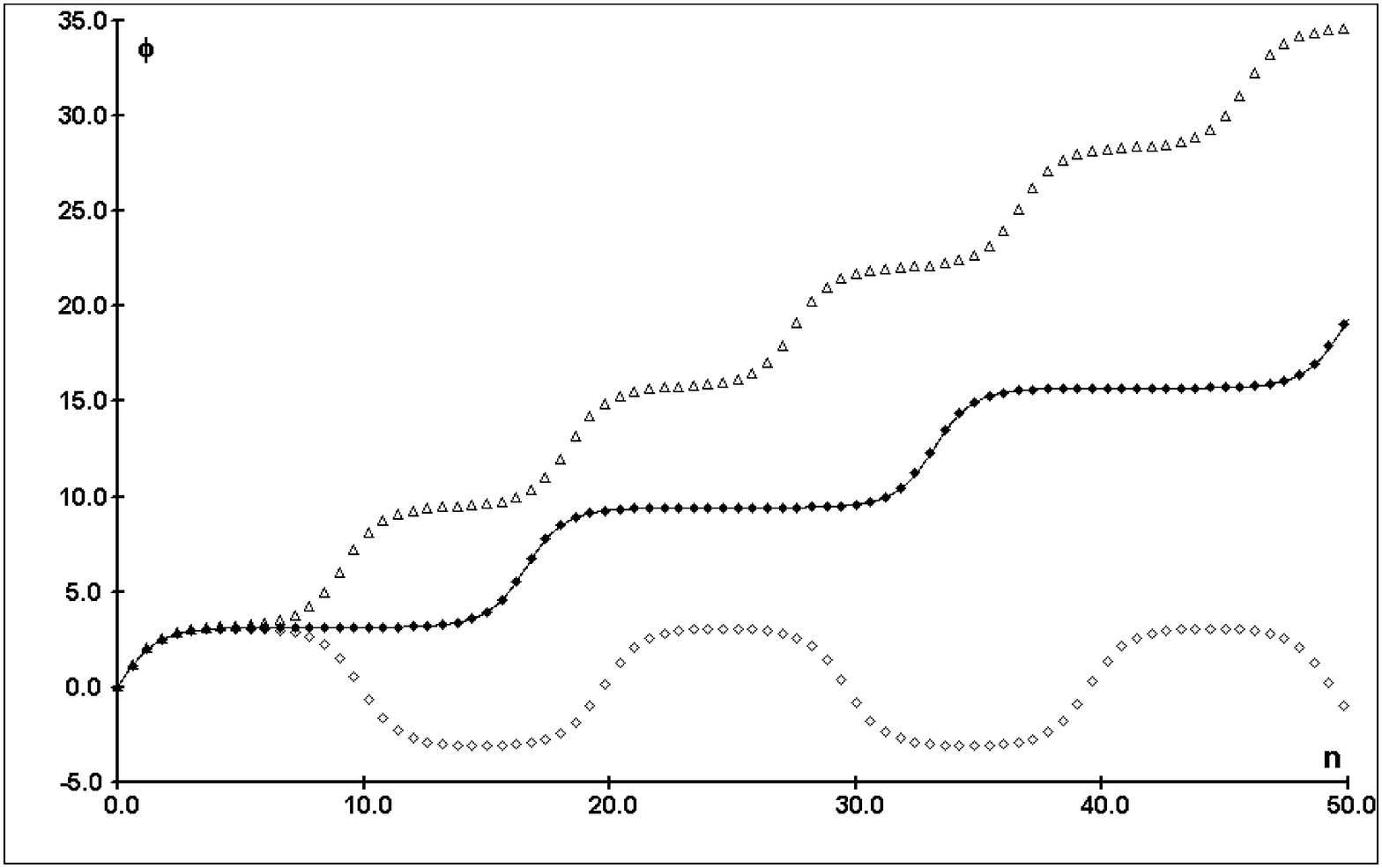} \par
\end{figure}

In the close neighbourhood of the separatrix 
the modified discrete gradient scheme behaves similarly to the discrete gradient method and 
its qualitative behaviour is perfect. What is more, 
also the quantitative results are very good (compare Table~\ref{error-sep}). 
Fig.~\ref{2E-06} compares the behaviour of our method with the leap-frog and 
implicit midpoint schemes for $p_0 = 2.000001$. The points generated by the 
modified discrete gradient method practically coincide with the exact solution 
(the relative error of the period is 0.59\%), almost as good result as that given by 
the discrete gradient scheme (the error is 0.25\%). The leap-frog 
scheme produces good qualitative behaviour but with the period two times smaller than 
the exact one. The implicit midpoint scheme gives wrong qualitative result: oscillations 
instead of rotation.

\section{Conclusions}

All methods considered in this paper are characterized by very high stability 
of periodic motions they generate (provided that $\ep$ is not too large). 
The average period is practically constant (with the accuracy 
close to $10^{-7}$ or  better) for a very long time (we checked even several millions 
of periods). 
The period and the amplitude perform regular small 
oscillations (they are relatively larger for both projection methods). The periodic 
character of 
these oscillations turns out to be of a systematic origin and we explained it  
considering rational approximations (with possibly small denominators) 
of the real number $T/\ep$. 
 
The main aim of this paper was the comparison of 
several numerical schemes. 
The standard leap-frog method, although non-integrable, is quite good when compared with
typical integrable discretizations. Its performance should be enhanced by use of projection
methods which impose the conservation of the energy integral. The projections work very well
for small values of the time step (e.g., the symmetric projection gives excellent results 
simulating the motion along the separatrix), 
while for larger time steps they produce relatively large
fluctuations of the period and the amplitude. In any case the projections produce much more accurate 
values of the average amplitude. The average period is of the same order, or even worse  
(in comparison to the standard leap-frog method). 

Surprising resonances occur for 
$p_0 \approx 1.21$ (for the leap-frog method) and $p_0 = 1.6$ (for the implicit midpoint rule). 
In the neighbourhood of these ``resonance'' values these methods have exclusively high 
accuracy of the estimated period (practically for any $\ep$), 
much better than all other methods. It would be interesting to explain this phenomenon. 

Discretizations found by Suris \cite{Sur89} are very stable but, 
in the same time, they have relatively 
large errors as compared to other numerical schemes. This is surprising because 
these methods are both integrable and symplectic. 
In this case the error (i.e., deviation from the exact solution) seems 
to be of a systematic origin.  
 We plan to construct appropriate modifications of Suris' discretizations 
in order to enhance their precision without destroying their stability.  

The discrete gradient method is (for any $\ep$ and any $p_0$) among the most accurate methods.
For rotating motions this is certainly the best method. 
We proposed a modification of the discrete gradient method which proved to be quite 
successful, especially when applied to simulate oscillating motions. Our new method is extremaly 
efficient for small oscillations. The relative error of the period computed by this method 
is less at least by 4 order of magnitude in comparison with other numerical schemes.

{\it Acknowledgements.}  The authors are grateful to Prof. Grzegorz Sitarski for useful comments
and turning our attention on Refs.~\cite{BFRB},\cite{GBB}. The first author was partially supported
by the Polish Ministry of Science and Higher Education
(grant no.\ 1 P03B 017 28).

\pagebreak

\clearpage

\renewcommand{\arraystretch}{1.6}

\begin{table}  
\caption{\footnotesize Stability of the period. Minimal, maximal and average values of 
$T_N$ for $p_0 = 1.95$, $\ep = 0.2$ ($T_{th} = 11.65758528$). }
\label{stability}

{\tiny
\begin{tabular}{|l|c|c|c|c|c|c|} \hline  discretization
 & \mc{2}{|c|}{maximal $T_N$} & \mc{2}{|c|}{minimal $T_N$} & \mc{2}{|c|}{average: 
${\bar T}_{avg} (N,100,200)$} \\ 
\cline{2-7}   
 &  $N < 100$ & $N \approx 1.8 \cdot 10^6$  & $N < 100$ & $N \approx 1.8 \cdot 10^6$  & $N = 0$ 
& $N = 1.8 \cdot 10^6$   \\ \hline 
leap-frog & 11.93166041 & 11.93166040 & 11.93164145 & 11.93164140 &  11.93165174 
& 11.93165162 \\ \hline  
Suris1 & 11.88885008 & 11.88885008 & 11.88883061 & 11.88883061 & 11.88884005 & 11.88884001 \\ \hline  
discrete gradient & 11.64698500 & 11.64698540 & 11.64697157 & 11.64697190 & 11.64697732 & 11.64697764 \\ \hline
\end{tabular}
}
\end{table}

\begin{table}  
\caption{\footnotesize Relative error of the amplitude ($A = A_{avg} (0,50)$).  \vspace{2ex} }
\label{error-amp}
{ \tiny 

\begin{tabular}{|l|r|r|r|r|r|r|r|r|} \hline  
\mc{1}{|c|}{$p_0$} & \mc{1}{|c|}{leap-frog} & \mc{1}{|c|}{Suris1} & \mc{1}{|c|}{Suris2} & \mc{1}{|c|}{gradient} &  
\mc{1}{|c|}{mod.\ grad. }  & \mc{1}{|c|}{projection}   & \mc{1}{|c|}{sym.\ proj.} & \mc{1}{|c|}{midpoint} \\ \hline 
\mc{9}{|c|}{ $\ep = 0.02$ } \\ \hline 
0.05 &5.00E-05	&1.50E-04	&1.00E-04	&-1.86E-08	&-1.87E-08  &-1.68E-08	&-1.71E-08 &-2.89E-08 \\ \hline
0.1	& 5.00E-05	&1.50E-04	&9.99E-05	&-1.85E-08	&-1.84E-08	&-1.10E-08	&-1.21E-08	&-6.01E-08 \\ \hline
0.3	&5.00E-05	&1.48E-04	&9.92E-05	&-1.74E-08	&-1.68E-08	&4.58E-08	&5.09E-08	&-3.95E-07 \\ \hline
0.5	&5.00E-05	&1.46E-04	&9.79E-05	&-1.55E-08	&-1.56E-08	&1.69E-08	&-8.59E-09	&-1.08E-06 \\ \hline
0.8	&5.02E-05	&1.39E-04	&9.47E-05	&-9.03E-09	&-8.37E-09	&-1.46E-07	&-2.05E-07	&-2.84E-06 \\ \hline
1.2	&5.13E-05	&1.26E-04	&8.86E-05	&-3.85E-09	&-3.88E-09	&-4.55E-09	&2.18E-09	&-7.00E-06 \\ \hline
1.6	&5.66E-05	&1.08E-04	&8.24E-05	&2.71E-09	&2.68E-09	&5.42E-10	&1.33E-08	&-1.53E-05 \\ \hline
1.8	&6.73E-05	&1.02E-04	&8.48E-05	&4.07E-09	&3.96E-09	&4.56E-09	&4.10E-09	&-2.49E-05 \\ \hline
\mc{9}{|c|}{ $\ep = 0.5$ } \\  \hline 
0.05	&2.54E-02	&8.52E-02	&5.58E-02	&-6.34E-03	&-6.87E-03	&-3.15E-02	&-3.16E-02	&-6.35E-03 \\ \hline
0.1	&2.55E-02	&8.52E-02	&5.57E-02	&-6.32E-03	&-6.80E-03	&-3.05E-02	&-3.14E-02	&-6.34E-03 \\ \hline
0.3	&2.58E-02	&8.46E-02	&5.56E-02	&-6.07E-03	&-6.59E-03	&-2.55E-02	&-2.76E-02	&-6.27E-03 \\ \hline
0.5	&2.65E-02	&8.34E-02	&5.54E-02	&-5.72E-03	&-6.13E-03	&-1.44E-02	&-2.13E-02	&-6.30E-03 \\ \hline
0.8	&2.81E-02	&8.05E-02	&5.46E-02	&-4.58E-03	&-5.01E-03	&8.86E-03	&-5.54E-03	&-6.12E-03 \\ \hline
1.2	&3.07E-02	&7.42E-02	&5.32E-02	&-2.44E-03	&-2.69E-03	&3.40E-02	&1.95E-02	&-6.54E-03 \\ \hline
1.6	&3.84E-02	&6.52E-02	&5.19E-02	&1.80E-04	&1.46E-04	&9.31E-03	&1.76E-02	&-9.00E-03 \\ \hline
1.8	&4.76E-02	&6.14E-02	&5.46E-02	&1.22E-03	&1.31E-03	&5.56E-03	&5.70E-03	&-1.36E-02 \\ \hline
\end{tabular}
}
\end{table}

\begin{table}  
\caption{\footnotesize Relative error of the period ($T = {\bar T}_{avg}(0,100,200)$). \vspace{2ex} }
\label{error-per}
{\tiny

\begin{tabular}{|l|r|r|r|r|r|r|r|r|} \hline  
\mc{1}{|c|}{$p_0$} & \mc{1}{|c|}{leap-frog} & \mc{1}{|c|}{Suris1} & \mc{1}{|c|}{Suris2} & \mc{1}{|c|}{gradient} &  
\mc{1}{|c|}{mod.\ grad. }  & \mc{1}{|c|}{projection}   & \mc{1}{|c|}{sym.\ proj.} & \mc{1}{|c|}{midpoint} \\ \hline 
\mc{9}{|c|}{ $\ep = 0.02$ } \\ \hline 
0.02	&-1.67E-05	&8.33E-05	&3.33E-05	&3.33E-05	&-3.34E-09	&-1.66E-05	&-1.66E-05	&3.33E-05 \\ \hline
0.05	&-1.66E-05	&8.33E-05	&3.33E-05	&3.33E-05	&-2.08E-08	&-1.64E-05	&-1.65E-05	&3.33E-05 \\ \hline
0.1	&-1.66E-05	&8.32E-05	&3.33E-05	&3.32E-05	&-8.34E-08	&-1.56E-05	&-1.59E-05	&3.32E-05 \\ \hline
0.3	&-1.59E-05	&8.18E-05	&3.30E-05	&3.26E-05	&-7.52E-07	&-6.74E-06	&-1.01E-05	&3.24E-05 \\ \hline
0.5	&-1.45E-05	&7.92E-05	&3.23E-05	&3.12E-05	&-2.10E-06	&1.11E-05	&1.70E-06	&3.07E-05 \\ \hline
0.8	&-1.08E-05	&7.28E-05	&3.10E-05	&2.79E-05	&-5.45E-06	&5.58E-05	&3.16E-05	&2.63E-05 \\ \hline
1.0	&-6.99E-06	&6.71E-05	&3.01E-05	&2.47E-05	&-8.63E-06	&9.86E-05	&6.05E-05	&2.20E-05 \\ \hline
1.2	&-1.48E-06	&6.05E-05	&2.95E-05	&2.07E-05	&-1.27E-05	&1.53E-04	&9.80E-05	&1.62E-05 \\ \hline
1.4	&6.86E-06	&5.37E-05	&3.03E-05	&1.56E-05	&-1.77E-05	&2.21E-04	&1.46E-04	&8.30E-06 \\ \hline
1.6	&2.12E-05	&4.92E-05	&3.52E-05	&9.31E-06	&-2.40E-05	&3.05E-04	&2.07E-04	&-3.63E-06 \\ \hline
1.8	&5.64E-05	&5.91E-05	&5.77E-05	&9.19E-07	&-3.24E-05	&4.08E-04	&2.87E-04	&-2.75E-05 \\ \hline
1.95	&2.17E-04	&1.90E-04	&2.03E-04	&-9.09E-06	&-4.24E-05	&4.99E-04	&3.72E-04	&-1.15E-04 \\ \hline
2.05	&-2.44E-04	&-2.78E-04	&-2.61E-04	&-1.14E-05	&-4.47E-05	&-3.47E-06	&1.57E-04	&1.14E-04 \\ \hline
2.2	&-9.25E-05	&-1.13E-04	&-1.03E-04	&-7.02E-06	&-4.04E-05	&-7.22E-05	&1.04E-04	&4.10E-05 \\ \hline 
2.5	&-5.71E-05	&-6.97E-05	&-6.34E-05	&-4.20E-06	&-3.75E-05	&-1.08E-04	&6.77E-05	&2.54E-05 \\ \hline 
3	&-4.45E-05	&-5.18E-05	&-4.81E-05	&-2.44E-06	&-3.58E-05	&-1.28E-04	&4.18E-05	&2.04E-05 \\ \hline
5	&-3.61E-05	&-3.83E-05	&-3.72E-05	&-7.26E-07	&-3.41E-05	&-1.43E-04	&1.46E-05	&1.75E-05  \\ \hline
\mc{9}{|c|}{ $\ep = 0.5$ } \\  \hline 
0.02	&-1.06E-02	&5.07E-02	&2.05E-02	&2.05E-02	&-2.03E-06	&-1.06E-02	&-1.07E-02	&2.05E-02 \\ \hline
0.05	&-1.06E-02	&5.07E-02	&2.05E-02	&2.05E-02	&-1.25E-05	&-1.05E-02	&-1.06E-02	&2.05E-02 \\ \hline
0.1	&-1.06E-02	&5.06E-02	&2.05E-02	&2.04E-02	&-5.02E-05	&-9.86E-03	&-1.03E-02	&2.04E-02 \\ \hline
0.3	&-1.01E-02	&4.97E-02	&2.03E-02	&2.01E-02	&-4.53E-04	&-3.29E-03	&-7.54E-03	&1.99E-02 \\ \hline
0.5	&-9.17E-03	&4.80E-02	&1.98E-02	&1.93E-02	&-1.27E-03	&1.01E-02	&-1.69E-03	&1.89E-02 \\ \hline
0.8	&-6.71E-03	&4.40E-02	&1.89E-02	&1.73E-02	&-3.30E-03	&4.43E-02	&1.42E-02	&1.64E-02 \\ \hline
1	&-4.13E-03	&4.02E-02	&1.83E-02	&1.53E-02	&-5.25E-03	&7.85E-02	&3.12E-02	&1.38E-02 \\ \hline
1.2	&-4.05E-04	&3.58E-02	&1.79E-02	&1.29E-02	&-7.74E-03	&1.24E-01	&5.55E-02	&1.03E-02 \\ \hline
1.4	&5.31E-03	&3.11E-02	&1.83E-02	&9.82E-03	&-1.09E-02	&1.84E-01	&9.04E-02	&5.56E-03 \\ \hline
1.6	&2.40E-02	&3.74E-02	&3.08E-02	&8.57E-03	&-2.13E-02	&4.11E-01	&2.14E-01	&-1.91E-03 \\ \hline
1.8	&4.28E-02	&3.27E-02	&3.80E-02	&6.42E-04	&-2.03E-02	&3.15E-01	&2.19E-01	&-1.56E-02 \\ \hline
1.95	&3.41E-01	&1.86E-01	&2.56E-01	&-5.72E-03	&-2.68E-02	&2.86E-01	&3.06E-01	&-5.94E-02 \\ \hline
2.05	&-1.17E-01	&-1.19E-01	&-1.19E-01	&-7.20E-03	&-2.83E-02	&-4.02E-02	&1.14E-01	&9.20E-02 \\ \hline
2.2	&-5.57E-02	&-5.95E-02	&-5.77E-02	&-4.45E-03	&-2.55E-02	&-4.42E-02	&8.08E-02	&2.70E-02 \\ \hline
2.5	&-3.68E-02	&-3.87E-02	&-3.77E-02	&-2.68E-03	&-2.37E-02	&-5.03E-02	&5.49E-02	&1.64E-02 \\ \hline
3	&-2.96E-02	&-2.96E-02	&-2.95E-02	&-1.57E-03	&-2.25E-02	&-5.51E-02	&3.34E-02	&1.34E-02 \\ \hline
5	&-2.68E-02	&-2.31E-02	&-2.50E-02	&-5.04E-04	&-2.14E-02	&-5.54E-02	&4.32E-03	&1.29E-02 \\ \hline
\end{tabular}
}
\end{table}

\begin{table}  
\caption{\footnotesize Relative error of the period in the neighbourhood of the separatrix. The blank space with a dot means that 
the qualitative behaviour of the discretization is wrong. \vspace{2ex} }
\label{error-sep}
{\tiny 

\begin{tabular}{|r|r|r|r|r|r|r|r|r|} \hline  
\mc{1}{|c|}{$p_0- 2.0$} & \mc{1}{|c|}{leap-frog} & \mc{1}{|c|}{Suris1} & \mc{1}{|c|}{Suris2} & \mc{1}{|c|}{gradient} &  
\mc{1}{|c|}{mod.\ grad. }  & \mc{1}{|c|}{projection}   & \mc{1}{|c|}{sym.\ proj.} & \mc{1}{|c|}{midpoint} \\ \hline 
\mc{9}{|c|}{ $\ep = 0.02$ } \\ \hline 
-1.0E-02 &	8.96E-04 &	8.50E-04 &	8.73E-04 &	-1.50E-05 &	-4.83E-05 &	5.19E-04 &	4.08E-04 &	-4.57E-04 \\ \hline
-1.0E-03 &	7.12E-03 &	7.06E-03 &	7.09E-03 &	-1.95E-05 &	-5.29E-05 &	5.14E-04 &	4.26E-04 &	-3.40E-03 \\ \hline
-1.0E-04 &	9.17E-02 &	9.16E-02 &	9.16E-02 &	-2.22E-05 &	-5.56E-05 &	5.05E-04 &	4.34E-04 &	-2.40E-02 \\ \hline
-1.0E-05 &	\mc{1}{|c|}{$\cdot$} &	\mc{1}{|c|}{$\cdot$} &	\mc{1}{|c|}{$\cdot$} &	-2.43E-05 &	-5.58E-05 &	4.99E-04 &	4.40E-04 &	-1.03E-01 \\ \hline
-1.0E-06 &	\mc{1}{|c|}{$\cdot$} &	\mc{1}{|c|}{$\cdot$} &	\mc{1}{|c|}{$\cdot$} &	-2.80E-05 &	-5.69E-05 &	4.94E-04 &	4.43E-04 &	-2.13E-01 \\ \hline
-1.0E-07 &	\mc{1}{|c|}{$\cdot$} &	\mc{1}{|c|}{$\cdot$} &	\mc{1}{|c|}{$\cdot$} &	-7.33E-05 &	-2.09E-05 &	4.91E-04 &	4.46E-04 &	-3.08E-01 \\ \hline
-1.0E-08 &	\mc{1}{|c|}{$\cdot$} &	\mc{1}{|c|}{$\cdot$} &	\mc{1}{|c|}{$\cdot$} &	 1.38E-04 &	 1.15E-04 &	4.88E-04 &	4.48E-04 &	-3.83E-01 \\ \hline
-1.0E-09 &	\mc{1}{|c|}{$\cdot$} &	\mc{1}{|c|}{$\cdot$} &	\mc{1}{|c|}{$\cdot$} &	-1.61E-03 &	 1.18E-03 &	4.86E-04 &	4.50E-04 &	-4.43E-01 \\ \hline
1.0E-08 &  -4.15E-01 &	-4.15E-01 &	-4.15E-01 &	-5.16E-05 &	-4.23E-06 &	2.82E-04 &	3.32E-04 &	\mc{1}{|c|}{$\cdot$}  \\ \hline
1.0E-07 &	-3.44E-01 &	-3.44E-01 &	-3.44E-01 &	-1.59E-05 &	-6.26E-05 &	2.60E-04 &	3.15E-04 &	\mc{1}{|c|}{$\cdot$}  \\ \hline
1.0E-06 &	-2.54E-01 &	-2.54E-01 &	-2.54E-01 &	-2.90E-05 &	-6.44E-05 &	2.31E-04 &	2.94E-04 &	\mc{1}{|c|}{$\cdot$}  \\ \hline
1.0E-04 &	-4.26E-02 &	-4.27E-02 &	-4.27E-02 &	-2.22E-05 &	-5.55E-05 &	8.93E-05 &	1.76E-04 &	3.38E-02  \\ \hline
1.0E-03 &	-6.68E-03 &	-6.73E-03 &	-6.70E-03 &	-1.96E-05 &	-5.29E-05 &	1.62E-04 &	2.69E-04 &	3.49E-03  \\ \hline
1.0E-01 &	-1.46E-04 &	-1.74E-04 &	-1.60E-04 &	-9.25E-06 &	-4.26E-05 &	-3.91E-05 &	1.31E-04 &	6.62E-05 \\ \hline
\mc{9}{|c|}{ $\ep = 0.5$ } \\  \hline 
-1.0E-02 &	\mc{1}{|c|}{$\cdot$} &	\mc{1}{|c|}{$\cdot$} &	\mc{1}{|c|}{$\cdot$} &	-9.51E-03 &	-3.06E-02 &	2.23E-01 &	3.31E-01 &	-1.54E-01 \\ \hline
-1.0E-03 &	\mc{1}{|c|}{$\cdot$} &	\mc{1}{|c|}{$\cdot$} &	\mc{1}{|c|}{$\cdot$} &	-1.24E-02 &	-3.36E-02 &	1.59E-01 &	3.31E-01 &	-3.21E-01 \\ \hline
-1.0E-04 &	\mc{1}{|c|}{$\cdot$} &	\mc{1}{|c|}{$\cdot$} &	\mc{1}{|c|}{$\cdot$} &	-1.41E-02 &	-3.53E-02 &	1.19E-01 &	3.26E-01 &	-4.48E-01 \\ \hline
-1.0E-05 &	\mc{1}{|c|}{$\cdot$} &	\mc{1}{|c|}{$\cdot$} &	\mc{1}{|c|}{$\cdot$} &	-1.52E-02 &	-3.65E-02 &	9.09E-02 &	3.22E-01 &	-5.36E-01 \\ \hline
-1.0E-06 &	\mc{1}{|c|}{$\cdot$} &	\mc{1}{|c|}{$\cdot$} &	\mc{1}{|c|}{$\cdot$} &	-1.61E-02 &	-3.73E-02 &	7.16E-02 &	3.19E-01 &	-6.01E-01 \\ \hline
-1.0E-07 &	\mc{1}{|c|}{$\cdot$} &	\mc{1}{|c|}{$\cdot$} &	\mc{1}{|c|}{$\cdot$} &	-1.67E-02 &	-3.80E-02 &	5.87E-02 &	3.17E-01 &	-6.49E-01 \\ \hline
-1.0E-08 &	\mc{1}{|c|}{$\cdot$} &	\mc{1}{|c|}{$\cdot$} &	\mc{1}{|c|}{$\cdot$} &	-1.73E-02 &	-3.86E-02 &	4.46E-02 &	3.16E-01 &	-6.88E-01 \\ \hline
-1.0E-09 &	\mc{1}{|c|}{$\cdot$} &	\mc{1}{|c|}{$\cdot$} &	\mc{1}{|c|}{$\cdot$} &	-1.73E-02 &	-3.86E-02 &	3.55E-02 &	3.14E-01 &	-7.18E-01 \\ \hline
1.0E-08 &	-7.24E-01 &	-7.22E-01 &	-7.23E-01 &	-1.72E-02 &	-3.85E-02 &	-5.68E-02 &	2.42E-01 &	\mc{1}{|c|}{$\cdot$}  \\ \hline
1.0E-07 &	-6.90E-01 &	-6.88E-01 &	-6.89E-01 &	-1.67E-02 &	-3.80E-02 &	-5.61E-02 &	2.35E-01 &	\mc{1}{|c|}{$\cdot$}  \\ \hline
1.0E-06 &	-6.47E-01 &	-6.45E-01 &	-6.46E-01 &	-1.61E-02 &	-3.73E-02 &	-5.53E-02 &	2.26E-01 &	\mc{1}{|c|}{$\cdot$}  \\ \hline
1.0E-04 &	-5.12E-01 &	-5.08E-01 &	-5.10E-01 &	-1.41E-02 &	-3.53E-02 &	-5.72E-02 &	1.96E-01 &	\mc{1}{|c|}{$\cdot$}  \\ \hline
1.0E-03 &	-3.97E-01 &	-3.93E-01 &	-3.96E-01 &	-1.24E-02 &	-3.36E-02 &	-4.36E-02 &	1.80E-01 &	\mc{1}{|c|}{$\cdot$}  \\ \hline
1.0E-01 &	-8.11E-02 &	-8.44E-02 &	-8.28E-02 &	-5.86E-03 &	-2.69E-02 &	-4.18E-02 &	9.80E-02 &	4.62E-02  \\ \hline
\end{tabular}
}
\end{table}

\end{document}